\documentclass[a4paper,UKenglish,cleveref, autoref]{lipics-v2019}
\pdfoutput=1
\nolinenumbers
\usepackage{mathtools} 
\usepackage{amstext} 
\usepackage{amssymb} 
\usepackage{bbold} 
\usepackage{amsthm} 

\usepackage[utf8]{inputenc} 

\usepackage{relsize} 
\usepackage{microtype} 
\usepackage{csquotes} 
\usepackage{ragged2e} 

\usepackage{hyperref} 
\usepackage[compress]{cite} 

\usepackage{graphicx} 
\usepackage[usenames,dvipsnames]{xcolor} 
\usepackage{stackengine}
\usepackage{tikz} 
\usepackage{circuitikz} 
\usetikzlibrary{
	arrows,
	shapes,
	decorations,
	intersections,
	backgrounds,
	positioning,
	circuits.ee.IEC
	}

\usepackage{tikzit}
\newcommand{\naturals}{\mathbb{N}}
\newcommand{\integers}{\mathbb{Z}}
\newcommand{\integersMod}[1]{\mathbb{Z}_{#1}}
\newcommand{\reals}{\mathbb{R}}
\newcommand{\complexs}{\mathbb{C}}
\newcommand{\nonstd}[1]{{^\star #1}}
\newcommand{\stdpartSym}{\operatorname{st}}
\newcommand{\stdpart}[1]{\stdpartSym\left(#1\right)}
\newcommand{\starNaturals}{\nonstd{\naturals}} 
\newcommand{\starIntegers}{\nonstd{\integers}} 
\newcommand{\starIntegersMod}[1]{\nonstd{\integersMod{#1}}}
\newcommand{\starComplexs}{\nonstd{\complexs}} 
\newcommand{\starReals}{\nonstd{\reals}} 
\newcommand{\suchthat}[2]{\left\{#1 \: \middle\vert \: #2\right\}}

\newcommand{\ket}[1]{\left| #1 \right\rangle}
\newcommand{\bra}[1]{\left\langle #1 \right|}
\newcommand{\braket}[2]{\left\langle #1 \middle| #2 \right\rangle}

\newcommand{\fHilb}{\operatorname{fHilb}}
\newcommand{\Hilb}{\operatorname{Hilb}}
\newcommand{\starHilb}{^\star\!\Hilb}

\newcommand{\obj}[1]{\operatorname{obj}}

\newcommand{\kstates}[1]{\operatorname{K}\left(#1\right)}

\tikzset{
  rectangle with rounded corners north west/.initial=4pt,
  rectangle with rounded corners south west/.initial=4pt,
  rectangle with rounded corners north east/.initial=4pt,
  rectangle with rounded corners south east/.initial=4pt,
}
\makeatletter
\pgfdeclareshape{rectangle with rounded corners}{
  \inheritsavedanchors[from=rectangle] 
  \inheritanchorborder[from=rectangle]
  \inheritanchor[from=rectangle]{center}
  \inheritanchor[from=rectangle]{north}
  \inheritanchor[from=rectangle]{south}
  \inheritanchor[from=rectangle]{west}
  \inheritanchor[from=rectangle]{east}
  \inheritanchor[from=rectangle]{north east}
  \inheritanchor[from=rectangle]{south east}
  \inheritanchor[from=rectangle]{north west}
  \inheritanchor[from=rectangle]{south west}
  \backgroundpath{
    \southwest \pgf@xa=\pgf@x \pgf@ya=\pgf@y
    \northeast \pgf@xb=\pgf@x \pgf@yb=\pgf@y
    \pgfkeysgetvalue{/tikz/rectangle with rounded corners north west}{\pgf@rectc}
    \pgfsetcornersarced{\pgfpoint{\pgf@rectc}{\pgf@rectc}}
    \pgfpathmoveto{\pgfpoint{\pgf@xa}{\pgf@ya}}
    \pgfpathlineto{\pgfpoint{\pgf@xa}{\pgf@yb}}
    \pgfkeysgetvalue{/tikz/rectangle with rounded corners north east}{\pgf@rectc}
    \pgfsetcornersarced{\pgfpoint{\pgf@rectc}{\pgf@rectc}}
    \pgfpathlineto{\pgfpoint{\pgf@xb}{\pgf@yb}}
    \pgfkeysgetvalue{/tikz/rectangle with rounded corners south east}{\pgf@rectc}
    \pgfsetcornersarced{\pgfpoint{\pgf@rectc}{\pgf@rectc}}
    \pgfpathlineto{\pgfpoint{\pgf@xb}{\pgf@ya}}
    \pgfkeysgetvalue{/tikz/rectangle with rounded corners south west}{\pgf@rectc}
    \pgfsetcornersarced{\pgfpoint{\pgf@rectc}{\pgf@rectc}}
    \pgfpathclose
 }
}
\makeatother

\tikzstyle{green dot}=[fill=YellowGreen!66!White, draw=black, shape=circle]
\tikzstyle{red dot}=[fill=Red, draw=black, shape=circle]
\tikzstyle{box}=[fill=white, draw=black, shape=rectangle, minimum width=1cm, minimum height=1cm]
\tikzstyle{small box}=[fill=white, draw=black, shape=rectangle, minimum width=0.3cm, minimum height=0.3cm]
\tikzstyle{wide box}=[fill=white, draw=black, shape=rectangle, minimum width=1.5cm, minimum height=1cm]
\tikzstyle{state} = [draw, rectangle with rounded corners,rectangle with rounded corners north west=0pt,rectangle with rounded corners south west=8pt,rectangle with rounded corners north east=0pt,rectangle with rounded corners south east=8pt,inner sep=2pt,minimum height=6mm,minimum width=6mm,fill=white]
\tikzstyle{effect} = [draw, rectangle with rounded corners,rectangle with rounded corners north west=8pt,rectangle with rounded corners south west=0pt,rectangle with rounded corners north east=8pt,rectangle with rounded corners south east=0pt,inner sep=2pt,minimum height=6mm,minimum width=6mm,fill=white]

\newcommand{\picturescaling}{0.8}
\newcommand{\picturescalingmore}{0.6}

\newcommand{\finElements}[1]{\left(#1\right)_\text{fin}}

\title{A Diagrammatic Approach to Quantum Dynamics}

\author{Stefano Gogioso}{University of Oxford}{stefano.gogioso@cs.ox.ac.uk}{https://orcid.org/0000-0001-7879-8145}{}

\Copyright{Stefano Gogioso} 



\acknowledgements{
	This publication was made possible through the support of a grant from the John Templeton Foundation. The opinions expressed in this publication are those of the authors and do not necessarily reflect the views of the John Templeton Foundation.
}

\titlerunning{A Diagrammatic Approach to Quantum Dynamics}
\authorrunning{S. Gogioso}

\EventEditors{Markus Roggenbach and Ana Sokolova}
\EventNoEds{2}
\EventLongTitle{8th Conference on Algebra and Coalgebra in Computer Science (CALCO 2019)}
\EventShortTitle{CALCO 2019}
\EventAcronym{CALCO}
\EventYear{2019}
\EventDate{June 3--6, 2019}
\EventLocation{London, United Kingdom}
\EventLogo{}
\SeriesVolume{139}
\ArticleNo{16}

\begin{document}
\maketitle

\begin{abstract}
	We present a diagrammatic approach to quantum dynamics based on the categorical algebraic structure of strongly complementary observables. We provide physical semantics to our approach in terms of quantum clocks and quantisation of time. We show that quantum dynamical systems arise naturally as the algebras of a certain dagger Frobenius monad, with the morphisms and tensor product of the category of algebras playing the role, respectively, of equivariant transformations and synchronised parallel composition of dynamical systems. We show that the Weyl Canonical Commutation Relations between time and energy are an incarnation of the bialgebra law and we derive Schr\"{o}dinger's equation from a process-theoretic perspective. Finally, we use diagrammatic symmetry-observable duality to prove Stone's proposition and von Neumann's Mean Ergodic proposition, recasting the results as two faces of the very same coin.
\end{abstract}

\section{Introduction}
\label{section:introduction}

It is hard to overstate the importance of quantum dynamics: in a very deep sense, it truly makes the world go round. In computer science, more specifically, it is the driving force behind the processes which underpin the entirety of quantum computation. Despite this crucial role, quantum dynamics is rarely considered directly in quantum information and in the foundations of quantum computing, being instead relegated to a lower level of abstraction. In this work we aim to change that, bringing dynamics on a par with information and circuits by developing a fully diagrammatic approach based on categorical algebra.

Our work fits within the framework of categorical quantum mechanics \cite{abramsky2004categorical,abramsky2009,coecke_kissinger_book} and uses the graphical calculus of string diagrams for symmetric monoidal categories \cite{selinger2011survey,joyal1991geometry}. We consider a particularly well-behaved kind of Hopf algebras/bialgebras---closely related to compact quantum groups \cite{woronowicz1987compact,woronowicz1998compact} and sometimes known as \emph{interacting quantum observables} \cite{Coecke2011interacting,duncan2016interacting}---arising as strongly complementary pairs of symmetric $\dagger$-Frobenius algebras \cite{vicary2011categorical,coecke2013new}. We show that the algebras of a certain dagger Frobenius monad \cite{heunen2016monads} correspond to quantum dynamical systems, that the morphisms between algebras correspond to equivariant transformations and that the natural tensor product in the category of algebras corresponds to synchronised parallel composition of dynamical systems. We further show that the dagger monoidal structure corresponds to symmetry-observable duality between time and energy, so that the Hamiltonian observables for quantum dynamical systems arise as the coalgebras of the corresponding dagger Frobenius comonad.

To showcase the expressive power of our formalism, we derive some cornerstone results of quantum dynamics in a diagrammatic fashion: the Weyl Canonical Commutation Relations, Schr\"{o}dinger's Equation, Stone's proposition and von Neumann's Mean Ergodic proposition. We use non-standard analysis \cite{gogioso2017infinite,gogioso2018qft} to deal with infinite-dimensional quantum systems.

\section{Interacting Quantum Observables}
\label{section:interacting-quantum-observables}

\subsection{Quantum Observables}

If $\mathcal{G}$ is an object in a dagger symmetric monoidal category, a \emph{$\dagger$-Frobenius algebra} $\raisebox{-0.5pt}{\hbox{\input{symbols/ZdotSym.tex}}}\!\!$ on $\mathcal{G}$ is a pair of a monoid $(\mathcal{G},\raisebox{-2.5pt}{\hbox{\input{symbols/ZmultSym.tex}}}\!,\!\raisebox{-2.5pt}{\hbox{\input{symbols/ZunitSym.tex}}}\!\!)$ and a comonoid $(\mathcal{G},\raisebox{-2.5pt}{\hbox{\input{symbols/ZcomultSym.tex}}}\!,\!\raisebox{-2.5pt}{\hbox{\input{symbols/ZcounitSym.tex}}}\!\!) := (\mathcal{G},\raisebox{-2.5pt}{\hbox{\input{symbols/ZmultSym.tex}}}\!^\dagger,\!\raisebox{-2.5pt}{\hbox{\input{symbols/ZunitSym.tex}}}\!\!^\dagger)$ related by \emph{Frobenius law}:
\begin{equation}
	\hfill
	\scalebox{\picturescaling}{$
		\input{figures/frobeniusLawL.tikz}
		\raisebox{0.75cm}{\hspace{2mm}$=$\hspace{2mm}}
		\input{figures/frobeniusLawM.tikz}
		\raisebox{0.75cm}{\hspace{2mm}$=$\hspace{2mm}}
		\input{figures/frobeniusLawR.tikz}
	$}
	\hfill
\end{equation}
A $\dagger$-Frobenius algebra is \emph{quasi-special} if the comultiplication is an isometry up to some invertible scalar $\xi$ and it is \emph{special} if it is an actual isometry (i.e. $\xi = 1$):
\begin{equation}
	\hfill
	\scalebox{\picturescaling}{$
		\input{figures/specialLawL.tikz}
		\raisebox{0.75cm}{\hspace{-3mm}$=$\hspace{4mm} $\xi^\dagger \xi$ \hspace{-6mm}}
		\begin{tikzpicture}
	\path [use as bounding box] (-1.00cm,-0.75cm) -- (1.00cm,-0.75cm) -- (1.00cm,0.75cm) -- (-1.00cm,0.75cm) -- (-1.00cm,-0.75cm);
	\begin{pgfonlayer}{nodelayer}
		\node (0) at (0.00,-0.75) {};
		\node (1) at (0.00,0.75) {};
	\end{pgfonlayer}
	\begin{pgfonlayer}{edgelayer}
		\draw [out=90.00,in=270.00] (0.center) to (1.center);
	\end{pgfonlayer}
\end{tikzpicture}

	$}
	\hfill
\end{equation}
The positive scalar $N_{\raisebox{-0.5pt}{\hbox{\input{symbols/ZdotSym.tex}}}\!\!} := \xi^\dagger \xi$ is known as the \emph{normalisation factor} for the algebra. Every quasi-special $\dagger$-Frobenius algebra is proportional to a special one, which we shall see shortly to have physical significance: we use quasi-special algebras merely for reasons of notational convenience.
The cup and cap induced by a $\dagger$-Frobenius algebra satisfy the \emph{snake equation}:
\begin{equation}
	\label{eqn_snakeEqn}
	\hfill
	\scalebox{\picturescaling}{$
		\input{figures/snakeL.tikz}
		\raisebox{0.75cm}{\hspace{2mm}$=$\hspace{-2mm}}
        \begin{tikzpicture}
	\path [use as bounding box] (-1.00cm,-0.75cm) -- (1.00cm,-0.75cm) -- (1.00cm,0.75cm) -- (-1.00cm,0.75cm) -- (-1.00cm,-0.75cm);
	\begin{pgfonlayer}{nodelayer}
		\node (0) at (0.00,-0.75) {};
		\node (1) at (0.00,0.75) {};
	\end{pgfonlayer}
	\begin{pgfonlayer}{edgelayer}
		\draw [out=90.00,in=270.00] (0.center) to (1.center);
	\end{pgfonlayer}
\end{tikzpicture}

        \raisebox{0.75cm}{\hspace{-2mm}$=$\hspace{2mm}}
		\input{figures/snakeR.tikz}
	$}
	\hfill
\end{equation}
A $\dagger$-Frobenius algebra is \emph{symmetric} if the cup and cap it induces are symmetric:
\begin{equation}
	\hfill
	\scalebox{\picturescaling}{$
		\input{figures/symmetryCapL.tikz}
		\raisebox{0.75cm}{\hspace{-3mm}$=$\hspace{-3mm}}
		\input{figures/symmetryCapR.tikz}
		\hspace{0cm}
		\input{figures/symmetryCupL.tikz}
		\raisebox{0.75cm}{\hspace{-3mm}$=$\hspace{-3mm}}
		\input{figures/symmetryCupR.tikz}
	$}
	\hfill
\end{equation}
Special symmetric $\dagger$-Frobenius algebras are of fundamental importance to quantum information, because they correspond exactly to quantum observables (more precisely, to finite-dimensional C*-algebras \cite{vicary2011categorical}). Special \emph{commutative} $\dagger$-Frobenius algebras in particular correspond to non-degenerate quantum observables, i.e. orthonormal bases. The basis vectors are exactly the \emph{classical states} $\ket{g}$ for the algebra:
\begin{equation}
\label{eqn_classicalStates}
	\hfill
	\scalebox{\picturescaling}{$
		\input{figures/classicalCopyL.tikz}
		\raisebox{1.25cm}{\hspace{-1mm}$=$\hspace{3mm}}
		\input{figures/classicalCopyR.tikz}
		\hspace{0.5cm}
		\input{figures/classicalDeleteL.tikz}
		\raisebox{1.25cm}{\hspace{-3mm}$=$\hspace{5mm} 1}
		\hspace{1.5cm}
		\raisebox{0.5cm}{
		\input{figures/classicalTransL.tikz}
		\raisebox{0.75cm}{\hspace{-1mm}$=$\hspace{2mm}}
		\input{figures/classicalTransR.tikz}
		}
	$}
	\hfill
\end{equation}
The leftmost two equations are the \emph{copying} and \emph{deleting} of the classical elements, while the rightmost equation says that classical elements are self-conjugate with respect to the observable. More generally, quasi-special commutative $\dagger$-Frobenius algebras $\raisebox{-0.5pt}{\hbox{\input{symbols/ZdotSym.tex}}}\!\!$ correspond to orthogonal bases in which all vectors have the same square norm $\braket{g}{g} = N_{\raisebox{-0.5pt}{\hbox{\input{symbols/ZdotSym.tex}}}\!\!}$. Because of this correspondence, we will henceforth liberally refer to quasi-special symmetric $\dagger$-Frobenius algebras as \emph{quantum observables}, as customary within the categorical quantum mechanics literature.\footnote{Though the term is usually referred to \emph{special} algebras, our generalisation to quasi-special maintains an exact correspondence with observables in $\fHilb$ and hence does not change the semantics.} We write $\kstates{\raisebox{-0.5pt}{\hbox{\input{symbols/ZdotSym.tex}}}\!\!}$ for the set of classical states.

\subsection{Interacting Quantum Observables}

\begin{definition}
	Let $(\mathcal{G},\raisebox{-0.5pt}{\hbox{\input{symbols/ZdotSym.tex}}}\!\!,\raisebox{-0.5pt}{\hbox{\input{symbols/XdotSym.tex}}}\!\!)$ be a pair of quasi-special symmetric $\dagger$-Frobenius algebras on the same system $\mathcal{G}$. We say that $(\mathcal{G},\raisebox{-0.5pt}{\hbox{\input{symbols/ZdotSym.tex}}}\!\!,\raisebox{-0.5pt}{\hbox{\input{symbols/XdotSym.tex}}}\!\!)$ is a \emph{strongly complementary pair}---or a pair of \emph{interacting} quantum observables---if the following equations---the \emph{bialgebra law}, the two \emph{coherence laws} and the \emph{bone law}---are satisfied:
	\begin{equation}
		\hfill
		\scalebox{\picturescaling}{$
			\input{figures/bialgebraL.tikz}
			\raisebox{1cm}{\hspace{4mm}$=$\hspace{-1mm}}
			\input{figures/bialgebraR.tikz}
			\hspace{0.8cm}
			\input{figures/greenDeleteL.tikz}
			\raisebox{0.75cm}{\hspace{0mm}$=$\hspace{1mm}}
			\input{figures/greenDeleteR.tikz}
			\hspace{0.8cm}
			\input{figures/greenCopyL.tikz}
			\raisebox{0.75cm}{\hspace{0mm}$=$\hspace{1mm}}
			\input{figures/greenCopyR.tikz}
			\hspace{0.8cm}
			\input{figures/boneLawL.tikz}
			\raisebox{0.65cm}{\hspace{0mm}$=$\hspace{5mm} $1$}
		$}
		\hfill
	\end{equation}
	As part of the definition, we also require that the unitary map $\!\!\raisebox{-2.5pt}{\hbox{\input{symbols/antipodeSym.tex}}}\!\!\!: \mathcal{G} \rightarrow \mathcal{G}$ defined below---the \emph{antipode}---is self-adjoint (or equivalently self-inverse):
	\begin{equation}
		\hfill
		\scalebox{\picturescaling}{$
			\input{figures/antipodeL.tikz}
			\raisebox{0.75cm}{\hspace{-1mm}$:=$\hspace{5mm}}
			\input{figures/antipodeC.tikz}
			\raisebox{0.75cm}{\hspace{5mm}$=$\hspace{5mm}}
			\input{figures/antipodeR.tikz}
		$}
		\hfill
	\end{equation}
\end{definition}
Interacting quantum observables automatically satisfy \emph{Hopf's law} and are thus examples of Hopf algebras:
\begin{equation}
	\hfill
	\scalebox{\picturescaling}{$
		\input{figures/hopfLawL.tikz}
		\raisebox{1cm}{\hspace{0mm}$=$\hspace{-2mm}}
		\input{figures/hopfLawC.tikz}
		\raisebox{1cm}{\hspace{-2mm}$=$\hspace{0mm}}
		\input{figures/hopfLawR.tikz}
	$}
	\hfill
\end{equation}
Because we are working in a dagger symmetric monoidal category, all equations above imply their dagger versions, so that $(\mathcal{G},\raisebox{-0.5pt}{\hbox{\input{symbols/ZdotSym.tex}}}\!\!,\raisebox{-0.5pt}{\hbox{\input{symbols/XdotSym.tex}}}\!\!)$ is a pair of interacting quantum observables if and only if $(\mathcal{G},\raisebox{-0.5pt}{\hbox{\input{symbols/XdotSym.tex}}}\!\!,\raisebox{-0.5pt}{\hbox{\input{symbols/ZdotSym.tex}}}\!\!)$ is. To get some intuition as to the meaning of the defining equations, we look at the characterisation \cite{gogioso2017mermin,kissinger2012thesis} of the pairs with $\raisebox{-0.5pt}{\hbox{\input{symbols/ZdotSym.tex}}}\!\!$ special and commutative in the category $\fHilb$ of finite-dimensional Hilbert spaces and complex linear maps.
\begin{proposition}
	The pairs $(\mathcal{G},\raisebox{-0.5pt}{\hbox{\input{symbols/ZdotSym.tex}}}\!\!,\raisebox{-0.5pt}{\hbox{\input{symbols/XdotSym.tex}}}\!\!)$ of interacting quantum observables in $\fHilb$ where $\raisebox{-0.5pt}{\hbox{\input{symbols/ZdotSym.tex}}}\!\!$ is special and commutative are exactly the \emph{group algebras} $\complexs[G]$ for finite groups $G$, with $\raisebox{-0.5pt}{\hbox{\input{symbols/ZdotSym.tex}}}\!\!$ corresponding to some orthonormal basis $(\ket{g})_{g \in G}$ labelled by the group elements, $\raisebox{-2.5pt}{\hbox{\input{symbols/XunitSym.tex}}}\! := \ket{1_{G}}$ and $\raisebox{-2.5pt}{\hbox{\input{symbols/XmultSym.tex}}}\!$ the linear extension of the multiplication of $G$:
	\begin{equation}
		\hfill
		\scalebox{\picturescaling}{$
			\input{figures/groupMultL.tikz}
			\raisebox{1.25cm}{\hspace{4mm}$=$\hspace{0mm}}
			\input{figures/groupMultR.tikz}
		$}
		\hfill
	\end{equation}
	Then antipode always corresponds to the group inverse:
	\begin{equation}
		\hfill
		\scalebox{\picturescaling}{$
			\input{figures/groupInvL.tikz}
			\raisebox{1cm}{\hspace{0mm}$=$\hspace{0mm}}
			\input{figures/groupInvR.tikz}
		$}
		\hfill
	\end{equation}
	The classical states $\ket{\chi}$ for $G$ are exactly those in the following form, where $\chi: G \rightarrow \complexs$ is a multiplicative character for $G$:
	\begin{equation}
		\hfill
		\ket{\chi} := \sum_{g \in G} \chi(g)\ket{g}
		\hfill
	\end{equation}
	If $G$ is abelian, then the pair $(\mathcal{G},\raisebox{-0.5pt}{\hbox{\input{symbols/XdotSym.tex}}}\!\!,\raisebox{-0.5pt}{\hbox{\input{symbols/ZdotSym.tex}}}\!\!)$ corresponds to the Pontryagin dual group $G^\wedge$.
\end{proposition}
By thinking of the comonoid $(\raisebox{-2.5pt}{\hbox{\input{symbols/ZcomultSym.tex}}}\!,\!\raisebox{-2.5pt}{\hbox{\input{symbols/ZcounitSym.tex}}}\!\!)$ as copying/deleting group elements and of the monoid $(\raisebox{-2.5pt}{\hbox{\input{symbols/XmultSym.tex}}}\!,\raisebox{-2.5pt}{\hbox{\input{symbols/XunitSym.tex}}}\!)$ as a ``coherent''/linear version of the group multiplication and unit, the defining equations for interacting quantum observables acquire a rather obvious meaning:
\begin{enumerate}
	\item the bialgebra law and left coherence law say that group multiplication sends elements which can be copied/deleted to elements which can be copied/deleted;
	\item the right coherence law and the bone law say that the group unit is itself and element which can be copied/deleted;
	\item the requirement that the antipode is self-inverse is necessary to prove Hopf's law, which in turn implies that elements have inverses (given by the antipode itself);
	\item the requirement that the antipode is self-inverse also implies that group multiplication sends self-conjugate elements for $\raisebox{-0.5pt}{\hbox{\input{symbols/ZdotSym.tex}}}\!\!$ to self-conjugate elements for $\raisebox{-0.5pt}{\hbox{\input{symbols/ZdotSym.tex}}}\!\!$ and the the group unit is itself self-conjugate for $\raisebox{-0.5pt}{\hbox{\input{symbols/ZdotSym.tex}}}\!\!$ \cite{duncan2016interacting,gogioso2017mermin}.
\end{enumerate}
This means that we can think of interacting quantum observables as ``coherent groups'', i.e. groups embedded into an environment in which $\dagger$-Frobenius algebras are available (e.g. in the compact-closed complex linear context). This is consistent with the fact that such coherent groups are particularly well-behaved examples of compact quantum groups \cite{woronowicz1987compact,woronowicz1998compact,gogioso2017thesis}---at least within the context of dagger compact categories of finite-dimensional vectors spaces over a field equipped with a self-inverse automorphism (acting as \emph{conjugation}).

\subsection{Dagger Frobenius Monads}

Given any monoid $(\mathcal{G},\raisebox{-2.5pt}{\hbox{\input{symbols/XmultSym.tex}}}\!,\raisebox{-2.5pt}{\hbox{\input{symbols/XunitSym.tex}}}\!)$ in a monoidal category, we can always define a monad by sending $\mathcal{H} \mapsto \mathcal{H} \otimes \mathcal{G}$, $f \mapsto f \otimes \operatorname{id}_{\mathcal{G}}$ and considering the following multiplication and unit:
\begin{equation}
	\hfill
	\scalebox{\picturescaling}{$
		\raisebox{0.75cm}{$\mu_{\mathcal{H}}$\hspace{3mm}$:=$\hspace{1mm}}
		\input{figures/monadMult.tikz}
		\hspace{3cm}
		\raisebox{0.75cm}{$\eta_{\mathcal{H}}$\hspace{3mm}$:=$\hspace{1mm}}
		\input{figures/monadUnit.tikz}
	$}
	\hfill
\end{equation}
The monad laws reduce to associativity and bilateral unitality for the monoid itself:
\begin{equation}
	\hfill
	\scalebox{\picturescaling}{$
		\input{figures/monadSquareLawL.tikz}
		\raisebox{0.75cm}{\hspace{1mm}$=$\hspace{1mm}}
		\input{figures/monadSquareLawR.tikz}
		\hspace{1.5cm}
		\input{figures/monadTriangleLaw1L.tikz}
		\raisebox{0.75cm}{\hspace{1mm}$=$\hspace{2mm}}
		\input{figures/monadTriangleLaw1R.tikz}
		\hspace{1.5cm}
		\input{figures/monadTriangleLaw2L.tikz}
		\raisebox{0.75cm}{\hspace{2mm}$=$\hspace{2mm}}
		\input{figures/monadTriangleLaw2R.tikz}
	$}
	\hfill
\end{equation}
If the category is dagger monoidal and the monoid is part of a $\dagger$-Frobenius algebra, then the monad is in fact a \emph{dagger Frobenius monad} \cite{heunen2016monads}, i.e. it satisfies the following law relating it to the comonad induced by the comonoid $(\mathcal{G},\raisebox{-2.5pt}{\hbox{\input{symbols/XcomultSym.tex}}}\!,\raisebox{-2.5pt}{\hbox{\input{symbols/XcounitSym.tex}}}\!)$:
\begin{equation}
	\hfill
	\scalebox{\picturescaling}{$
		\input{figures/monadFrobeniusLawL.tikz}
		\raisebox{0.75cm}{\hspace{2mm}$=$\hspace{2mm}}
		\input{figures/monadFrobeniusLawR.tikz}
	$}
	\hfill
\end{equation}
Because no ambiguity can arise, we write $\_ \otimes \raisebox{-0.5pt}{\hbox{\input{symbols/XdotSym.tex}}}\!\!$ to denote both the monad given by $(\mathcal{G},\raisebox{-2.5pt}{\hbox{\input{symbols/XmultSym.tex}}}\!,\raisebox{-2.5pt}{\hbox{\input{symbols/XunitSym.tex}}}\!)$ and the comonad given by $(\mathcal{G},\raisebox{-2.5pt}{\hbox{\input{symbols/XcomultSym.tex}}}\!,\raisebox{-2.5pt}{\hbox{\input{symbols/XcounitSym.tex}}}\!)$. The \emph{algebras} for the dagger Frobenius monad $\_ \otimes \raisebox{-0.5pt}{\hbox{\input{symbols/XdotSym.tex}}}\!\!$ are the morphisms $\alpha: \mathcal{H} \otimes \mathcal{G} \rightarrow \mathcal{H}$ such that:
\begin{equation}
	\hspace{-7mm}
	\scalebox{\picturescalingmore}{$
		\input{figures/algebraMultL.tikz}
		\raisebox{2cm}{\hspace{-2mm}$=$\hspace{0mm}}
		\input{figures/algebraMultR.tikz}
		\hspace{0cm}
		\input{figures/algebraUnitL.tikz}
		\raisebox{2cm}{\hspace{-2mm}$=$\hspace{-10mm}}
		\begin{tikzpicture}
	\path [use as bounding box] (-2.00cm,-2.00cm) -- (2.00cm,-2.00cm) -- (2.00cm,2.00cm) -- (-2.00cm,2.00cm) -- (-2.00cm,-2.00cm);
	\begin{pgfonlayer}{nodelayer}
		\node (0) at (0.00,-2.00) {};
		\node (1) at (0.00,2.00) {};
	\end{pgfonlayer}
	\begin{pgfonlayer}{edgelayer}
		\draw [out=90.00,in=270.00] (0.center) to (1.center);
	\end{pgfonlayer}
\end{tikzpicture}

		\hspace{0cm}
		\input{figures/algebraTransL.tikz}
		\raisebox{2cm}{\hspace{0mm}$=$\hspace{-4mm}}
		\input{figures/algebraTransR.tikz}
	$}
\end{equation}
In the original \cite{heunen2016monads}, these are referred to more specifically as \emph{FEM-algebras}, for \emph{Frobenius Eilenberg-Moore algebras}. Just as the ``Eilenberg-Moore/EM'' qualifier is customarily dropped for algebras of a monad, we shall here also drop the ``Frobenius/FEM'' qualifier for the algebras of dagger Frobenius monads, because FEM-algebras are the natural notion in the dagger Frobenius context.
\begin{proposition}
	\label{proposition:dagger-frobenius-monad-transpose-condition}
	When $(\mathcal{G},\raisebox{-0.5pt}{\hbox{\input{symbols/ZdotSym.tex}}}\!\!,\raisebox{-0.5pt}{\hbox{\input{symbols/XdotSym.tex}}}\!\!)$ is a pair of interacting quantum observables, the rightmost condition above in the definition of algebras for the monad $\_ \otimes \raisebox{-0.5pt}{\hbox{\input{symbols/XdotSym.tex}}}\!\!$ can be equivalently reformulated as follows:
	\begin{equation}
		\hfill
		\scalebox{\picturescaling}{$
			\input{figures/algebraTransGreenL.tikz}
			\raisebox{1.5cm}{\hspace{5mm}$=$\hspace{5mm}}
			\input{figures/algebraTransGreenR.tikz}
		$}
		\hfill
	\end{equation}
\end{proposition}

\section{Quantum Clocks}
\label{section:quantum-clocks}

In a very practical sense, time is ticked by clocks. If we know that a dynamical system and a clock are synchronised, then we can know the exact state of the system without ever looking at it, by just knowing what time the clock is displaying (assuming we know the initial state for the system). The kinds of dynamics admissible for systems synchronised with a given clock depend on the structure of the clock: clocks with more time states can be synchronised with more systems. If one interprets dynamics as time-translation symmetry---the approach that we take in this work---this means that the dynamical system has to be a representation for whatever time-translation group is associated with the clock.

The interpretation of dynamics as time-translation symmetry may appear causally problematic: after all, what does it mean to have entanglement across time? A full discussion of this issue would take more words than can fit in the margins of these pages, but we hope the following will at least convince the reader that the point of view we've adopted might not be as preposterous as it may at first seem.

When thinking about time, we can take two perspectives: an \emph{internal} perspective---reflecting time as \emph{dynamically} experienced by those immersed in its flow---and an \emph{external} perspective---reflecting time as \emph{statically} experienced by those staring at it from the outside. Mathematically, the internal perspective roughly corresponds to the idea of time evolution as the solution of differential equations: an instantaneous state is given and propagated forward instant by instant according to the laws of dynamics. The external perspective, on the other hand, corresponds to the idea of spacetime: everything which happens in the entire history of a mathematical objects is already crystallised in front of the eyes of those studying it, its evolution merely a matter of choosing and relating equal-time slices.

In the case of quantum dynamics, the external perspective amounts to thinking of dynamical systems as static, their entire history of evolution $(\ket{\psi_t})_t$ encoded in an entangled state $\sum_{t} \ket{\psi_t} \otimes \ket{t}$, where we selected a suitably large ancillary quantum system $\mathcal{G}$---a \emph{quantum clock}, as we shall call it---equipped with a choice of \emph{time observable} $\raisebox{-0.5pt}{\hbox{\input{symbols/ZdotSym.tex}}}\!\!$ labelling the time states $\ket{t}$. This idea was already clear in the formulation of Feynman's clock \cite{feynman1982simulating,feynman1986quantum}: therein, the computation of the dynamics of a quantum system is reduced to the computation of the ground state of a certain Hamiltonian, resulting precisely in the system-clock entangled state detailed above. Causality and dynamics arise from the choice of a specific group structure on the time states---the \emph{time-translation group}---corresponding to a strongly complementary quantum observable $\raisebox{-0.5pt}{\hbox{\input{symbols/XdotSym.tex}}}\!\!$: the time observable chooses the time slices for the dynamical system, the group structure relates them causally and dynamically.

For an internal observer, living inside the quantum dynamical system, the classical evolution $(\ket{\psi_t})_t$ is indistinguishable from the external application of time-translation symmetry to slices of definite time value $\ket{t}$: to such an internal observer, time behaves as an external classical parameter. To an external observer, on the other hand, the difference between the two perspectives is truly one of expressive power: taking the internal perspective forces them to work in the time observable for the quantum clock, while taking the external perspective allows the to freely change their point of view, yielding additional insights and more direct proofs of canonical results.

\subsection{Quantum clocks from interacting quantum observables -- Take I}

Here, we restrict our attention to four inter-related kinds of dynamics: discrete periodic, discrete, continuous periodic and continuous.
\begin{enumerate}
	\item Systems with continuous dynamics correspond to the the usual choice of time-translation group $\reals$ (continuous time).
	\item Systems with continuous periodic dynamics correspond to a choice of time-translation group in the form $\reals/T\integers$ for some positive period $T \in \reals$. They are exactly the systems with continuous dynamics where the dynamics are periodic.
	\item Systems with discrete dynamics correspond to the choice of time-translation group $\integers$. Sampling systems with continuous dynamics at equally spaced discrete intervals of time yields systems with discrete dynamics. These are the systems which are effectively synchronised with ordinary clocks (as long as we assume access to infinite time counters which never cycle).
	\item Systems with discrete periodic dynamics correspond to a choice of time-translation group in the form $\integersMod{N}$ for some positive period $N \in \naturals$. Sampling systems with continuous periodic dynamics (with period $T \in \reals$) at equally spaced discrete intervals of time (spaced by some positive $\Delta t \in \reals$ dividing the period $T$) yields systems with discrete dynamics (with $N := T / \Delta t$). These are the systems which are effectively synchronised with ordinary clocks having finite time counters---from 12-hour wall clocks, with their 43200 time states, to high-precision atomic clocks.
\end{enumerate}
When $G$ is one of the time-translation groups above, a dynamical system \emph{governed by} $G$---i.e. one which can be synchronised with clocks having $G$ as the associated time-translation group---is simply a representation $(U_g)_{g \in G}$ of $G$ in some appropriate category modelling the physical context of interest. In particular, a quantum dynamical system is just a unitary representation $(U_g)_{g \in G}$ of $G$ on some Hilbert space $\mathcal{H}$, with appropriate continuity requirements impose where necessary.

The identification of quantum dynamical systems with unitary representations is the mainstream view in quantum dynamics, but it introduces an unpleasant asymmetry between the physical systems---which are quantum---and time---which is instead a classical parameter external to the quantum realm. This asymmetry did not escape the attention of the founders of quantum mechanics, and the history of attempts to quantise time is long and rife with controversy. We refer the reader interested in such history to some very good works dedicated specifically to the topic \cite{Hilgevoord2005time,Pashby2015time,roberts2012time,butterfield2014time}: in this work, we will instead avoid such controversies altogether, by taking an external ``static'' view of quantum dynamical systems.

For the purposes of dynamics, we have seen that a clock---a physical system---can be abstracted to a time-translation group $G$---an algebraic structure that it can be endowed with. If we take an ordinary clock---the states of which are the possible instants of the time ticks---and we quantise it, we obtain a \emph{quantum clock}---the states of which are now \emph{wavefunctions} over the space of states for the original clock. In this interpretation, quantum clocks should be quantum systems equipped with the structure of a group algebra $\complexs[G]$ for the time-translation group $G$.

In the case $G = \integersMod{N}$ of discrete periodic dynamics we already know what to do: a quantum clock is finite-dimensional Hilbert space $\mathcal{G}$---a quantum system, living in the dagger compact category $\fHilb$---endowed with a pair $(\mathcal{G},\raisebox{-0.5pt}{\hbox{\input{symbols/ZdotSym.tex}}}\!\!,\raisebox{-0.5pt}{\hbox{\input{symbols/XdotSym.tex}}}\!\!)$ of interacting quantum observables corresponding to $\complexs[\integersMod{N}]$---a pair of categorical algebraic structures. Unlike classical clocks---where a single algebraic object was needed---quantum clocks need an interacting \emph{pair} of algebraic structures: one to pin down the time states (the observable $\raisebox{-0.5pt}{\hbox{\input{symbols/ZdotSym.tex}}}\!\!$) and another one to endow them with the $\integersMod{N}$ group structure (the observable $\raisebox{-0.5pt}{\hbox{\input{symbols/XdotSym.tex}}}\!\!$).

\subsection{Infinite-dimensional quantum systems}

This approach---modelling quantum clocks using interacting quantum observables---will work well for finite-dimensional quantum dynamical systems with discrete periodic dynamics, which are by themselves of significant interest: they were extensively studied by Weyl \cite{weyl1927quantenmechanik,weyl1950theory} and can be used to formalise Feynman's clock construction \cite{feynman1982simulating,feynman1986quantum,mcclean2013feynman}. However, it cannot immediately be generalised to the other kinds of dynamics which we are interested in: the only pairs of interacting quantum observables in the dagger symmetric monoidal category $\Hilb$ of Hilbert spaces and continuous linear maps are the ones corresponding to finite groups. Technically, $\Hilb$ doesn't even have quantum observables as we defined them: an orthonormal basis $(\ket{e_i})_{i = 1}^{\infty}$ of an infinite-dimensional separable Hilbert space is still associated with pair of a commutative comultiplication $\raisebox{-2.5pt}{\hbox{\input{symbols/ZcomultSym.tex}}}\! := \sum_{i=1}^{\infty} (\ket{e_i} \otimes \ket{e_i}) \bra{e_i}$ and multiplication $\raisebox{-2.5pt}{\hbox{\input{symbols/ZmultSym.tex}}}\! = \raisebox{-2.5pt}{\hbox{\input{symbols/ZcomultSym.tex}}}\!^\dagger$ satisfying the Frobenius law---as well as an alternative equation making classical states self-conjugate---but we have to let go of the counit $\!\raisebox{-2.5pt}{\hbox{\input{symbols/ZcounitSym.tex}}}\!\!$ and unit $\!\raisebox{-2.5pt}{\hbox{\input{symbols/ZunitSym.tex}}}\!\!$ in the passage from finite- to infinite-dimensional Hilbert spaces \cite{abramsky2012hstar}. Indeed, the unit for such an algebra would have to take the form $\!\raisebox{-2.5pt}{\hbox{\input{symbols/ZunitSym.tex}}}\!\! = \sum_{i=1}^{\infty} \ket{e_i}$, a vector which would have infinite norm.

In order to gain access to interacting quantum observables corresponding to infinite group algebras, we work in a symmetric monoidal category of non-standard Hilbert spaces and $\starComplexs$-linear maps known as $\starHilb$ \cite{gogioso2017infinite,gogioso2018qft}, where $\starComplexs$ is the field of non-standard complex numbers. The objects of $\starHilb$ are non-standard Hilbert spaces which are \emph{hyperfinite}-dimensional, i.e. which have orthonormal bases in the form $(\ket{e_i})_{i=1}^{n}$ where $n \in \starNaturals$ is a non-standard natural number (the \emph{dimension} of the non-standard Hilbert space); in particular, $\starHilb$ contains symmetric monoidal sub-categories equivalent to $\fHilb$ and $\Hilb$ (up to infinitesimals). Even though $n$ may be an infinite natural, from a non-standard perspective the objects of $\starHilb$ are finite-dimensional spaces: this means that $\starHilb$ is dagger compact, has special symmetric $\dagger$-Frobenius algebras and contains the extra strongly complementary pairs which we need to talk about quantum dynamics.

The easiest way to work with $\starHilb$ is by using the \emph{Transfer Principle}: if a construction indexed by $n$ can be made on $n$-dimensional Hilbert spaces for all $n \in \naturals$, then it can be uniquely extended to $n$-dimensional non-standard Hilbert spaces for all $n \in \starNaturals$. For an extensive introduction to non-standard analysis and the Transfer Principle we refer the reader to Refs. \cite{robinson1974nonstandard,goldblatt1998hyperreals}. For example, if $(\ket{e_i})_{i=1}^{n}$ is an orthonormal basis for an $n$-dimensional Hilbert space, we can always define the unit for the associated quantum observable as $\!\raisebox{-2.5pt}{\hbox{\input{symbols/ZunitSym.tex}}}\!\! = \sum_{i=1}^{n} \ket{e_i}$: by the Transfer Principle, this means that we can also do so for an orthonormal basis $(\ket{e_i})_{i=1}^{n}$ of an object of $\starHilb$ where $n$ is an infinite natural. The vector $\!\raisebox{-2.5pt}{\hbox{\input{symbols/ZunitSym.tex}}}\!\! = \sum_{i=1}^{n} \ket{e_i}$ has a well-defined infinite square norm $\sum_{i=1}^n \braket{e_i}{e_i} = n$ and can be normalised to $\frac{1}{\sqrt{n}}\sum_{i=1}^{n} \ket{e_i}$ as one would ordinarily do in a finite-dimensional Hilbert space. This latter example shows that, when $n$ is infinite, $\starHilb$ features some genuinely new quantum states: $\frac{1}{\sqrt{n}}\sum_{i=1}^{n} \ket{e_i}$ is \emph{finite}, in the sense that it has finite norm, but not \emph{near-standard}, in the sense that it is not infinitesimally close to any vector in the corresponding standard Hilbert space. These extra states are the key to constructing the interacting quantum observables we need.

The trick to constructing quantum clocks in $\starHilb$ is to think of all time-translation groups as actually discrete and periodic, at least in the non-standard sense. Consider the abelian group $\starIntegersMod{\omega}$ formed by the non-standard integers modulo some positive non-standard natural $\omega \in \starNaturals$. When $\omega$ is finite, these are the usual finite cyclic groups. When $\omega$ is infinite, however, these groups are always very large, and contain $\integers$ as a subgroup. Indeed, we can take the following representatives for the elements of $\starIntegersMod{\omega}$:
\begin{equation}
	\hfill
	\starIntegersMod{\omega} := \left(\left\{-\left\lfloor \frac{\omega-1}{2} \right\rfloor,...,+\left\lfloor \frac{\omega}{2} \right\rfloor\right\}, +,0\right)
	\hfill
\end{equation}
If $i,j \in \integers$ then $i+j$ is always finite and no modular reduction ever occurs, so that addition of $i$ and $j$ in $\starIntegersMod{\omega}$ is the same as addition in $\integers$.
Now let $\omega_\text{uv},\omega_\text{ir} \in \starReals$ be non-infinitesimal positive non-standard reals with $\omega_\text{uv}\omega_\text{ir} = \omega \in \starNaturals$ and consider the following subset of $\starReals$:
\begin{equation}
	\hfill
	\frac{1}{\omega_\text{uv}}\starIntegersMod{\omega} := \suchthat{\frac{n}{\omega_\text{uv}} \in \starReals}{n \in \left\{-\left\lfloor \frac{\omega-1}{2} \right\rfloor,...,+\left\lfloor \frac{\omega}{2} \right\rfloor\right\}}
	\hfill
\end{equation}
The subset $\frac{1}{\omega_\text{uv}}\starIntegersMod{\omega}$ inherits the group structure of $\starIntegersMod{\omega}$. The uv/ir suffixes for the numbers $\omega_\text{uv}$ and $\omega_\text{ir}$ originate from a habit, typical of quantum field theory, to distinguish between ``infra-red'' infinities--- arising because space is infinitely large---and ``ultra-violet'' infinities---arising because space is infinitely fine: the parameter $\omega_\text{uv}$ controls how fine the subdivision of $\starReals$ specified by $\frac{1}{\omega_\text{uv}}\starIntegersMod{\omega}$ is, while the parameter $\omega_\text{ir} = \omega / \omega_\text{uv}$ controls how large a portion of $\starReals$ it covers.

For different choices of parameters $\omega_\text{uv},\omega_\text{ir} \in \starReals$, the discrete periodic non-standard groups $\frac{1}{\omega_\text{uv}}\starIntegersMod{\omega}$ can be used to approximate all the time-translation groups which we are interested in. In what follows, we write $\finElements{\frac{1}{\omega_\text{uv}}\starIntegersMod{\omega}}$ for the subgroup formed by the finite elements, i.e. by those elements which are finite reals.
\footnote{Note to the reader versed in non-standard analysis: this is an \emph{external} subgroup and is only used for the purpose of connecting the non-standard groups to their standard counterparts. It is never used in any constructions within $\starHilb$.}
If $x \in \finElements{\frac{1}{\omega_\text{uv}}\starIntegersMod{\omega}}$, we write $\stdpart{x}$ for the unique standard real which is infinitesimally close to $x$.
\begin{proposition}
	\label{proposition:non-standard-time-translation-groups}
	Let $\omega_\text{uv},\omega_\text{ir} \in \starReals$ be non-infinitesimal positive non-standard reals such that $\omega:= \omega_\text{uv}\omega_\text{ir} \in \starNaturals$ is integer. The time-translation groups $G$ for discrete periodic, discrete, continuous periodic and continuous dynamics are exactly the standard groups which can be obtained as quotient by infinitesimals of the subgroup of finite elements of $\frac{1}{\omega_\text{uv}}\starIntegersMod{\omega}$:
	\begin{equation}
		\hfill
		G = \stdpart{\finElements{\frac{1}{\omega_\text{uv}}\starIntegersMod{\omega}}}
		\hfill
	\end{equation}
	More specifically, we have the following combinations:
	\begin{enumerate}
		\item if $\omega_\text{uv}$ is finite and $\omega_\text{ir}$ is finite we obtain the discrete periodic case $G = \frac{1}{\stdpart{\omega_\text{uv}}}\integersMod{\omega} \cong \integersMod{\omega}$;
		\item if $\omega_\text{uv}$ is finite and $\omega_\text{ir}$ infinite we obtain the discrete case $\frac{1}{\stdpart{\omega_\text{uv}}}\integers \cong \integers$;
		\item if $\omega_\text{uv}$ is infinite and $\omega_\text{ir}$ is finite we obtain the continuous periodic case $\reals/\stdpart{\omega_\text{ir}}\integers$;
		\item if $\omega_\text{uv}$ is infinite and $\omega_\text{ir}$ is infinite we obtain the continuous case $\reals$.
	\end{enumerate}
	Hence all standard time-translation groups listed above can be approximated, up to infinitesimals, by the subgroup of finite elements of a discrete periodic non-standard group.
\end{proposition}

\subsection{Quantum clocks from interacting quantum observables -- Take II}

Armed with our dagger compact category $\starHilb$ and with approximations of our favourite time-translation groups by discrete periodic non-standard groups, we are finally in a position to define our quantum clocks.

\begin{definition}
	A \emph{quantum clock}  is a pair of interacting quantum observables $(\mathcal{G},\raisebox{-0.5pt}{\hbox{\input{symbols/ZdotSym.tex}}}\!\!,\raisebox{-0.5pt}{\hbox{\input{symbols/XdotSym.tex}}}\!\!)$ in $\starHilb$ with $\raisebox{-0.5pt}{\hbox{\input{symbols/ZdotSym.tex}}}\!\!$ special commutative, equipped with a group isomorphism $(\kstates{\raisebox{-0.5pt}{\hbox{\input{symbols/ZdotSym.tex}}}\!\!},\raisebox{-2.5pt}{\hbox{\input{symbols/XmultSym.tex}}}\!,\raisebox{-2.5pt}{\hbox{\input{symbols/XunitSym.tex}}}\!) \cong \frac{1}{\omega_\text{uv}}\starIntegersMod{\omega}$ for some non-infinitesimal positive $\omega_\text{uv},\omega_\text{ir} \in \starReals$ with $\omega := \omega_\text{uv} \omega_\text{ir} \in \starNaturals$. We refer to $\frac{1}{\omega_\text{uv}}\starIntegersMod{\omega}$ as the \emph{time-translation group} and to $\stdpart{\finElements{\frac{1}{\omega_\text{uv}}\starIntegersMod{\omega}}}$ as the associated \emph{standard time-translation group}. We refer to the classical states $\kstates{\raisebox{-0.5pt}{\hbox{\input{symbols/ZdotSym.tex}}}\!\!}$ of $\raisebox{-0.5pt}{\hbox{\input{symbols/ZdotSym.tex}}}\!\!$ as the \emph{(clock) time states}---which we index as $\ket{t}$ using the elements $t \in \frac{1}{\omega_\text{uv}}\starIntegersMod{\omega}$---and to the observable $\raisebox{-0.5pt}{\hbox{\input{symbols/ZdotSym.tex}}}\!\!$ as the \emph{clock time observable}.
\end{definition}
Note that the specific choice of $\frac{1}{\omega_\text{uv}}\starIntegersMod{\omega}$---i.e. the specific choice of parameters $\omega_\text{uv},\omega_\text{ir}$---is part of the data of a quantum clock. We will only mention it explicitly when relevant, to lighten the notation.

\begin{proposition}
	\label{proposition:quantum-clocks-existence}
	Quantum clocks with non-standard time-translation group $\frac{1}{\omega_\text{uv}}\starIntegersMod{\omega}$ exist for all non-infinitesimal positive $\omega_\text{uv},\omega_\text{ir} \in \starReals$ with $\omega := \omega_\text{uv} \omega_\text{ir} \in \starNaturals$.
\end{proposition}

In the case of infinite-dimensional quantum clocks, working in the non-standard setting gives us access to a lot of states and linear maps which would not be well-defined in the standard setting, let alone continuous. On a quantum clock with time-translation group $\frac{1}{\omega_\text{uv}}\starIntegersMod{\omega}$, for example, we can construct the following \emph{plane-wave states} indexed by all $E \in \frac{1}{\omega_\text{ir}}\starIntegersMod{\omega}$ (note the switch from $\omega_\text{uv}$ to $\omega_\text{ir}$):
\begin{equation}
	\label{equation:clock-energy-states}
	\hfill
	\ket{E} := \sum_{t \in \frac{1}{\omega_\text{uv}}\starIntegersMod{\omega}} e^{i2\pi \frac{Et}{\omega}}\ket{t}
	\hfill
\end{equation}
In particular, we have that $\braket{E}{t} = e^{-i2\pi Et}$. This is exactly the phase that an energy eigenstate with energy $E$ acquires after time $t$ has passed. As the following result shows, this is no coincidence: in a quantum clock $(\mathcal{G},\raisebox{-0.5pt}{\hbox{\input{symbols/ZdotSym.tex}}}\!\!,\raisebox{-0.5pt}{\hbox{\input{symbols/XdotSym.tex}}}\!\!)$, the classical states $\ket{E}$ for $\raisebox{-0.5pt}{\hbox{\input{symbols/XdotSym.tex}}}\!\!$ always label the possible energy values that the corresponding dynamical systems can have.
\begin{proposition}
	\label{proposition:pontryagin-duality}
	In a quantum dynamical system with standard time-translation group $G$, the possible values for energy are always canonically labelled by the elements of the Pontryagin dual $G^\wedge$. If $(\mathcal{G},\raisebox{-0.5pt}{\hbox{\input{symbols/ZdotSym.tex}}}\!\!,\raisebox{-0.5pt}{\hbox{\input{symbols/XdotSym.tex}}}\!\!)$ is a quantum clock with time-translation group $\frac{1}{\omega_\text{uv}}\starIntegersMod{\omega}$, the classical states for $\raisebox{-0.5pt}{\hbox{\input{symbols/XdotSym.tex}}}\!\!$ are the plane-wave states of Equation \eqref{equation:clock-energy-states} and we have $(\kstates{\raisebox{-0.5pt}{\hbox{\input{symbols/XdotSym.tex}}}\!\!},\raisebox{-2.5pt}{\hbox{\input{symbols/ZmultSym.tex}}}\!,\!\raisebox{-2.5pt}{\hbox{\input{symbols/ZunitSym.tex}}}\!\!) \cong \frac{1}{\omega_\text{ir}}\starIntegersMod{\omega}$. If $G$ is the standard time-translation group associated to the quantum clock then:
	\begin{equation}
		\hfill
		G^\wedge = \stdpart{\finElements{\frac{1}{\omega_\text{ir}}\starIntegersMod{\omega}}}
		\hfill
	\end{equation}
	Hence the classical states of $\raisebox{-0.5pt}{\hbox{\input{symbols/XdotSym.tex}}}\!\!$ canonically label the possible energy levels for quantum dynamical systems that can be synchronised with the clock.
\end{proposition}

When $E$ and $t$ are both finite---i.e. when they have direct physical significance---we can manually check that $\stdpart{e^{-i2\pi Et}}$ yields the expected phase in the various models. In all four cases we have $\stdpart{e^{-i2\pi Et}} = e^{-i2\pi \stdpart{E}\stdpart{t}}$, with the domains of $\stdpart{t}$ and $\stdpart{E}$ ensuring that the expression is well-defined under all circumstances.
\begin{enumerate}
	\item For continuous dynamics, $\stdpart{t} \in \reals$ and $\stdpart{E} \in \reals$ and there are no issues.
	\item For continuous periodic dynamics, $\stdpart{t} \in \reals/T\integers$, so we need $\stdpart{E} \in \frac{1}{T}\integers$ for the phase to be well-defined. Indeed, $\frac{1}{T}\integers$ is the standard group we obtain when $\omega_\text{ir} = T$.
	\item For discrete dynamics, $\stdpart{t} \in \integers$ and values of $\stdpart{E}$ differing by $1$ will give the exact same phase: to have an exact correspondence, we therefore need $\stdpart{E} \in \reals/\integers$. Indeed, $\reals/\integers$ is the standard group we obtain when $\omega_\text{uv} = 1$.
	\item For discrete periodic dynamics, we have $\stdpart{t} \in \integersMod{\omega}$. This combines the requirements on $\stdpart{E}$ from both the previous cases: values of $\stdpart{E}$ differing by $1$ will correspond to the same phase, and the phase is only well-defined if $\stdpart{E}$ is divisible by $\omega$. Indeed, the standard group we obtain in this case ($\omega_\text{uv}=1$ and $\omega_\text{ir}=\omega$) is $\stdpart{E} \in \frac{1}{\omega}\integersMod{\omega}$.
\end{enumerate}
In light of the above, we adopt the following definition.
\begin{definition}
	Let $(\mathcal{G},\raisebox{-0.5pt}{\hbox{\input{symbols/ZdotSym.tex}}}\!\!,\raisebox{-0.5pt}{\hbox{\input{symbols/XdotSym.tex}}}\!\!)$ be a quantum clock. We refer to the classical states $\ket{E}$ of $\raisebox{-0.5pt}{\hbox{\input{symbols/XdotSym.tex}}}\!\!$ as \emph{clock energy states} and to the observable $\raisebox{-0.5pt}{\hbox{\input{symbols/XdotSym.tex}}}\!\!$ as the \emph{clock energy observable}.
\end{definition}

\section{Quantum Dynamical Systems}
\label{section:quantum-dynamical-systems}

In the previous Section, we have shown that certain pairs of interacting quantum observables in the dagger compact category $\starHilb$ can be used to model quantum clocks, i.e. quantum systems with additional structure singling out certain clock time states and the desired time-translation group structure on them. In this Section, we switch our attention to quantum dynamical systems.

\subsection{Quantum Dynamical Systems}

Let $(\mathcal{G},\raisebox{-0.5pt}{\hbox{\input{symbols/ZdotSym.tex}}}\!\!,\raisebox{-0.5pt}{\hbox{\input{symbols/XdotSym.tex}}}\!\!)$ be a quantum clock and consider an algebra $\alpha: \mathcal{H} \otimes \mathcal{G} \rightarrow \mathcal{H}$ for the dagger Frobenius monad $\_ \otimes \raisebox{-0.5pt}{\hbox{\input{symbols/XdotSym.tex}}}\!\!$. We look at the endomorphisms $\alpha_t: \mathcal{H} \rightarrow \mathcal{H}$ obtained by evaluating the algebra on clock time states $\ket{t}$:
\begin{equation}
	\hfill
	\scalebox{\picturescaling}{$
		\raisebox{1.5cm}{$\alpha_t$\hspace{6mm}$:=$\hspace{4mm}}
		\input{figures/alphaT.tikz}
	$}
	\hfill
\end{equation}
In terms of those endomorphisms, the defining equations for algebras take the following form:
\begin{equation}
	\hfill
	\scalebox{\picturescaling}{$
		\input{figures/alphaTmultL.tikz}
		\raisebox{1.5cm}{\hspace{2mm}$=$\hspace{2mm}}
		\input{figures/alphaTmultR.tikz}
		\hspace{1cm}
		\input{figures/alphaTunitL.tikz}
		\raisebox{1.5cm}{\hspace{2mm}$=$\hspace{-4mm}}
		\begin{tikzpicture}
	\path [use as bounding box] (-1.00cm,-1.50cm) -- (1.00cm,-1.50cm) -- (1.00cm,1.50cm) -- (-1.00cm,1.50cm) -- (-1.00cm,-1.50cm);
	\begin{pgfonlayer}{nodelayer}
		\node (0) at (0.00,-1.50) {};
		\node (1) at (0.00,1.50) {};
	\end{pgfonlayer}
	\begin{pgfonlayer}{edgelayer}
		\draw [out=90.00,in=270.00] (0.center) to (1.center);
	\end{pgfonlayer}
\end{tikzpicture}

		\hspace{1cm}
		\input{figures/alphaTtransL.tikz}
		\raisebox{1.5cm}{\hspace{2mm}$=$\hspace{2mm}}
		\input{figures/alphaTtransR.tikz}
	$}
	\hfill
\end{equation}
But these are exactly the equations defining a unitary representation $(\alpha_t)_t$ of the time-translation group! Clearly, we are off to a good start.

Algebras for the monad $\_ \otimes \raisebox{-0.5pt}{\hbox{\input{symbols/XdotSym.tex}}}\!\!$ form a category, with morphisms $\Phi: \alpha \rightarrow \beta$ from an algebra $\alpha: \mathcal{H} \otimes \mathcal{G} \rightarrow \mathcal{H}$ to another algebra $\beta: \mathcal{K} \otimes \mathcal{G} \rightarrow \mathcal{K}$ given by the linear maps $\Phi: \mathcal{H} \rightarrow \mathcal{K}$ which satisfy the following equation:
\begin{equation}
	\hfill
	\scalebox{\picturescaling}{$
		\input{figures/algebraMorphismL.tikz}
		\raisebox{1.75cm}{\hspace{6mm}$=$\hspace{6mm}}
		\input{figures/algebraMorphismR.tikz}
	$}
	\hfill
\end{equation}
Because the monad $\_ \otimes \raisebox{-0.5pt}{\hbox{\input{symbols/XdotSym.tex}}}\!\!$ is obtained from a monoid which is part of a pair $(\raisebox{-0.5pt}{\hbox{\input{symbols/ZdotSym.tex}}}\!\!,\raisebox{-0.5pt}{\hbox{\input{symbols/XdotSym.tex}}}\!\!)$ of interacting quantum observables with $\raisebox{-0.5pt}{\hbox{\input{symbols/ZdotSym.tex}}}\!\!$ commutative, the category of algebras has a symmetric monoidal structure, with the tensor product $\alpha \otimes \beta$ of two algebras defined as follows:
\begin{equation}
	\hfill
	\scalebox{\picturescaling}{$
		\input{figures/algebraTensorProduct.tikz}
	$}
	\hfill
\end{equation}
The following result shows that the category of algebras captures exactly quantum dynamical systems, equivariant maps between them and their natural notion of composition.
\begin{proposition}
	\label{proposition:algebras-dynamical-systems}
	Let $(\mathcal{G},\raisebox{-0.5pt}{\hbox{\input{symbols/ZdotSym.tex}}}\!\!,\raisebox{-0.5pt}{\hbox{\input{symbols/XdotSym.tex}}}\!\!)$ be a quantum clock. The algebras $\alpha$ for the dagger Frobenius monad $\_ \otimes \raisebox{-0.5pt}{\hbox{\input{symbols/XdotSym.tex}}}\!\!$ such that $\alpha_t$ is near-standard for all $t$ correspond to quantum dynamical systems for the standard time-translation, i.e. strongly continuous unitary representations $(\stdpart{\alpha_t})_{\stdpart{t} \in G}$ of the standard time-translation group. Morphisms between algebras correspond to equivariant maps for the representations. Tensor product of algebras corresponds to synchronised composition of quantum dynamical systems:
	\begin{equation}
		\hfill
		\scalebox{\picturescaling}{$
			\raisebox{2cm}{$(\alpha \otimes \beta)_t$\hspace{6mm}$:=$\hspace{4mm}}
			\input{figures/algebraTensorProductT.tikz}
            \raisebox{2cm}{\hspace{4mm}$=$\hspace{6mm} $\alpha_t \otimes \beta_t$}
		$}
		\hfill
	\end{equation}
\end{proposition}
\begin{definition}
	Let $(\mathcal{G},\raisebox{-0.5pt}{\hbox{\input{symbols/ZdotSym.tex}}}\!\!,\raisebox{-0.5pt}{\hbox{\input{symbols/XdotSym.tex}}}\!\!)$ be a quantum clock. A \emph{quantum dynamical system} for the quantum clock is an algebra for the dagger Frobenius monad $\_ \otimes \raisebox{-0.5pt}{\hbox{\input{symbols/XdotSym.tex}}}\!\!$. Morphisms of algebras will be referred to as \emph{equivariant maps} between quantum dynamical systems. Tensor product of algebras will be referred to as \emph{synchronised parallel composition} of quantum dynamical systems.
\end{definition}

\subsection{States and histories}

States of a system are a static concept. In a quantum dynamical system, we are instead more interested in the evolution of states under the dynamics:
	\begin{equation}
		\hfill
		\scalebox{\picturescaling}{$
			\input{figures/stateEvolution.tikz}
		$}
		\hfill
	\end{equation}
For a generic dynamical system $H$---e.g. seen as a topological space---the evolution of a state under the dynamics is usually written as a \emph{flow-line} $\Psi: \reals \rightarrow H$, a map from the time object to the dynamical system associating a state $\Psi(t) \in H$ to each instant point $t$ in time (with the obvious generalisation from continuous dynamics to the other three kinds). This is not, however, an exact correspondence in general: a given map $\reals \rightarrow H$ is often not going to be the flow-line of a state.

As we mentioned before, the traditional perspective on time in quantum dynamics is that time is an external classical parameter, so the definition of state evolution through flow-lines suffers from the issue described above. In our framework, on the other hand, ``time'' lives inside the same category as the quantum systems it governs, incarnated into the quantum clocks that tick it. We can exploit the additional algebraic structure available to show that flow-lines, realised inside the category of algebras, correspond exactly to the evolutions of states in quantum dynamical systems.

\begin{proposition}
	\label{proposition:histories}
	Let $(\mathcal{G},\raisebox{-0.5pt}{\hbox{\input{symbols/ZdotSym.tex}}}\!\!,\raisebox{-0.5pt}{\hbox{\input{symbols/XdotSym.tex}}}\!\!)$ be a quantum clock and $\alpha: \mathcal{H} \otimes \mathcal{G} \rightarrow \mathcal{H}$ be a quantum dynamical system for it. Then $\raisebox{-2.5pt}{\hbox{\input{symbols/XmultSym.tex}}}\!$ is also a quantum dynamical system for it---the quantum clock itself, governed by its own time. The morphisms of algebras $\Psi: \raisebox{-2.5pt}{\hbox{\input{symbols/XmultSym.tex}}}\! \rightarrow \alpha$: 
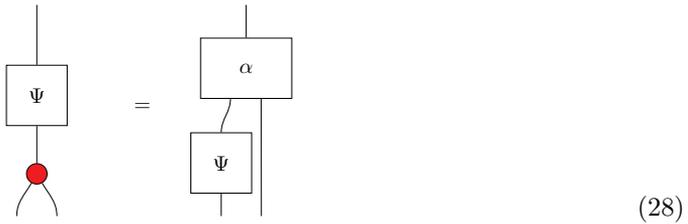
	\begin{equation}
		\hfill
		\scalebox{\picturescaling}{$
			\input{figures/historyL.tikz}
			\raisebox{1.75cm}{\hspace{6mm}$=$\hspace{6mm}}
			\input{figures/historyR.tikz}
		$}
		\hfill
	\end{equation}
	are exactly the evolutions of states of $\mathcal{H}$ under the dynamics of $\alpha$. We refer to such morphisms as the \emph{histories} of states.
\end{proposition}

\subsection{Hamiltonians}

Hamiltonians are often the very first concept that students of quantum dynamics are introduced to, so it may be surprising that we have not mentioned them so far. The reason for such a delay is that this work adopts a view of dynamics as time-translation symmetry, rather than as solution of certain differential equations: our objects of primary concern are unitary representations, not their infinitesimal generators. That said the \emph{Hamiltonian}---as the energy observable of a quantum dynamical system---is of paramount physical interest, so we now proceed to characterise it in our framework.

In previous sections, we have considered the dagger Frobenius monad $\_ \otimes \raisebox{-0.5pt}{\hbox{\input{symbols/XdotSym.tex}}}\!\!$ (with associated dagger Frobenius comonad $\_ \otimes \raisebox{-0.5pt}{\hbox{\input{symbols/XdotSym.tex}}}\!\!$) and we have seen that the algebras of $\_ \otimes \raisebox{-0.5pt}{\hbox{\input{symbols/XdotSym.tex}}}\!\!$ capture dynamics. We have also seen, when talking about quantum clocks, that the other quantum observable in the interacting pair, namely $\raisebox{-0.5pt}{\hbox{\input{symbols/ZdotSym.tex}}}\!\!$, is somehow associated with energy: one naturally wonders whether there is an algebraic connection between $\raisebox{-0.5pt}{\hbox{\input{symbols/ZdotSym.tex}}}\!\!$---which is dual to $\raisebox{-0.5pt}{\hbox{\input{symbols/XdotSym.tex}}}\!\!$---and Hamiltonians---which are dual to dynamics.

An alternative characterisation of a quantum observable is in terms of complete families of projectors. A complete family of projectors $(P_E: \mathcal{H} \rightarrow \mathcal{H})_{E \in X}$ is characterised by the following equations:
\begin{equation}
	\hfill
	\scalebox{\picturescaling}{$
		\input{figures/PEmultL.tikz}
		\raisebox{1.5cm}{\hspace{2mm}$=$\hspace{4mm}$\delta_{E,F}$ \hspace{0mm}}
		\input{figures/PEmultR.tikz}
		\hspace{1cm}
		\raisebox{1.5cm}{$\sum_E$ \hspace{0mm}}
		\input{figures/PEunitL.tikz}
		\raisebox{1.5cm}{\hspace{2mm}$=$\hspace{-4mm}}
		\begin{tikzpicture}
	\path [use as bounding box] (-1.00cm,-1.50cm) -- (1.00cm,-1.50cm) -- (1.00cm,1.50cm) -- (-1.00cm,1.50cm) -- (-1.00cm,-1.50cm);
	\begin{pgfonlayer}{nodelayer}
		\node (0) at (0.00,-1.50) {};
		\node (1) at (0.00,1.50) {};
	\end{pgfonlayer}
	\begin{pgfonlayer}{edgelayer}
		\draw [out=90.00,in=270.00] (0.center) to (1.center);
	\end{pgfonlayer}
\end{tikzpicture}

		\hspace{1cm}
		\input{figures/PEtransL.tikz}
		\raisebox{1.5cm}{\hspace{2mm}$=$\hspace{2mm}}
		\input{figures/PEtransR.tikz}
	$}
	\hfill
\end{equation}
If the labels for the projectors $P_E$ are taken from the clock energy states $E$, as necessary for the projectors associated with a Hamiltonian, then the equations above can be equivalently rewritten diagrammatically as the equations for coalgebras of the dagger Frobenius comonad $\_ \otimes \raisebox{-0.5pt}{\hbox{\input{symbols/ZdotSym.tex}}}\!\!$:
\begin{equation}
	\hspace{-7mm}
	\scalebox{\picturescalingmore}{$
		\input{figures/coalgebraMultL.tikz}
		\raisebox{2cm}{\hspace{0mm}$=$\hspace{1mm}}
		\input{figures/coalgebraMultR.tikz}
		\hspace{1cm}
		\input{figures/coalgebraUnitL.tikz}
		\raisebox{2cm}{\hspace{4mm}$=$\hspace{0mm}}
		\begin{tikzpicture}
	\path [use as bounding box] (-1.00cm,-2.00cm) -- (1.00cm,-2.00cm) -- (1.00cm,2.00cm) -- (-1.00cm,2.00cm) -- (-1.00cm,-2.00cm);
	\begin{pgfonlayer}{nodelayer}
		\node (0) at (0.00,-2.00) {};
		\node (1) at (0.00,2.00) {};
	\end{pgfonlayer}
	\begin{pgfonlayer}{edgelayer}
		\draw [out=90.00,in=270.00] (0.center) to (1.center);
	\end{pgfonlayer}
\end{tikzpicture}

		\hspace{1.5cm}
		\input{figures/coalgebraTransL.tikz}
		\raisebox{2cm}{\hspace{4mm}$=$\hspace{0mm}}
		\input{figures/coalgebraTransR.tikz}
	$}
\end{equation}
In the literature, these are also referred to as \emph{projector-valued spectra} \cite{coecke2007measurements}. We can give such coalgebras an operational interpretation as coherent versions of quantum measurements: if we feed a state $\ket{\psi}$ of $\mathcal{H}$ in input, we obtain in output an entangled state $\sum_{E} P_E \ket{\psi} \otimes \ket{E}$ of $\mathcal{H} \otimes \mathcal{G}$. Subsequently measuring $\mathcal{G}$ in the $(\ket{E})_E$ basis yields the usual von Neumann non-demolition measurement corresponding to the complete family of orthogonal projectors $(P_E)_E$: if outcome $E$ is observed, the state in $\mathcal{H}$ has collapsed to $P_E \ket{\psi}$.

Given a quantum dynamical system $\alpha$, we now show how to obtain the coalgebra for $\_ \otimes \raisebox{-0.5pt}{\hbox{\input{symbols/ZdotSym.tex}}}\!\!$ corresponding to its Hamiltonian. We will do so by proving Schr\"{o}dinger's Equation. In its differential version, Schr\"{o}dinger's Equation states that if $\ket{\psi_E}$ is an energy eigenstate with energy $E$ then the evolution of $\ket{\psi_E}$ in a quantum dynamical system $\alpha$ is given by the following equation:
\begin{equation}
	\hfill
	i \hbar \frac{d}{dt} \alpha_t \ket{\psi_E} = E \ket{\psi_E}
	\hfill
\end{equation}
The following exponentiated version of Schr\"{o}dinger's Equation provides the symmetry equivalent of the usual differential equation:
\begin{equation}
	\hfill
	\alpha_t \ket{\psi_E} = e^{-i 2\pi Et} \ket{\psi_E}
	\hfill
\end{equation}
where we have chosen energy and time units such that $h := 2 \pi \hbar = 1$.
\begin{proposition}
	\label{proposition:hamiltonian}
	Let $(\mathcal{G},\raisebox{-0.5pt}{\hbox{\input{symbols/ZdotSym.tex}}}\!\!,\raisebox{-0.5pt}{\hbox{\input{symbols/XdotSym.tex}}}\!\!)$ be a quantum clock and $\alpha: \mathcal{H} \otimes \mathcal{G} \rightarrow \mathcal{H}$ be a quantum dynamical system for it. Then $\alpha^\dagger: \mathcal{H} \rightarrow \mathcal{H} \otimes \mathcal{G}$ is a coalgebra for the dagger Frobenius comonad $\_ \otimes \raisebox{-0.5pt}{\hbox{\input{symbols/ZdotSym.tex}}}\!\!$, with projectors $P_E$ labelled by clock energy states $\ket{E}$:
	\begin{equation}
		\hfill
			\scalebox{\picturescaling}{$
				\raisebox{1.75cm}{$P_E$\hspace{8mm}$:=$\hspace{6mm}}
				\input{figures/PE.tikz}
				\raisebox{2.4cm}{\hspace{0mm}$\dfrac{1}{\omega}$}
			$}
		\hfill
	\end{equation}
	Note that $\frac{1}{\omega} = \frac{1}{N_{\raisebox{-0.5pt}{\hbox{\input{symbols/XdotSym.tex}}}\!\!}}$ is the normalisation factor for the state $\ket{E}$.
	The states invariant under projector $P_E$ satisfy Schr\"{o}dinger's Equation for energy $E$:
	\begin{equation}
		\hfill
			\scalebox{\picturescaling}{$
				\input{figures/SchrEq1L.tikz}
				\raisebox{2.3cm}{\hspace{5mm}$=$\hspace{5mm}}
				\input{figures/SchrEq1R.tikz}
				\raisebox{2.3cm}{\hspace{15mm}$\Rightarrow$\hspace{15mm}}
				\input{figures/SchrEq2L.tikz}
				\raisebox{2.3cm}{\hspace{5mm}$=$\hspace{5mm}}
				\input{figures/SchrEq2R.tikz}
			$}
		\hfill
	\end{equation}
	recalling that $\braket{E}{t} = e^{-i 2\pi Et }$. Hence the projectors $P_E$ are exactly the projectors onto the energy eigenspaces of the quantum dynamical system, so that $\alpha^\dagger$ is the projector-valued spectrum for the Hamiltonian observable.
\end{proposition}
\begin{definition}
	Let $(\mathcal{G},\raisebox{-0.5pt}{\hbox{\input{symbols/ZdotSym.tex}}}\!\!,\raisebox{-0.5pt}{\hbox{\input{symbols/XdotSym.tex}}}\!\!)$ be a quantum clock with time-translation group $\frac{1}{\omega_\text{uv}}\starIntegersMod{\omega}$ and let $\alpha: \mathcal{H} \otimes \mathcal{G} \rightarrow \mathcal{H}$ be a quantum dynamical system for it. The \emph{Hamiltonian} for $\alpha$ is the coalgebra $\alpha^\dagger$. The \emph{energy eigenstates} for $\alpha$ corresponding to clock energy $E \in \frac{1}{\omega_\text{ir}}\starIntegersMod{\omega}$ are the states $\ket{\psi}$ satisfying the following equation:
	\begin{equation}
		\hfill
			\scalebox{\picturescaling}{$
				\input{figures/eigenstateL.tikz}
				\raisebox{1.75cm}{\hspace{5mm}$=$\hspace{5mm}}
				\input{figures/eigenstateR.tikz}
			$}
		\hfill
	\end{equation}
\end{definition}
The simplicity and elegance of the characterisation given above for the Hamiltonian---the coalgebra obtained as adjoint of the algebra capturing the quantum dynamical system---showcases the power of the coherent approach we have adopted. By quantising clocks, the dual information about time/dynamics and energy is now held by the very same object: if we want to switch perspective, we only need to switch observable. This form of \emph{diagrammatic time/energy duality} will make it possible, in the coming section, to derive extremely compact diagrammatic proofs for some result of fundamental importance in quantum dynamics.

\section{Cornerstone Results}
\label{section:results}

In the previous Section, we have established a clear parallel between the language of quantum dynamics and the language of algebra. In this Section, we use that parallel to re-establish three cornerstone results of quantum dynamics in diagrammatic terms.

\subsection{Weyl Canonical Commutation Relations}

Traditionally, the Heisenberg Canonical Commutation Relations characterise the duality between position and momentum observables in the differential generators picture. The Weyl Canonical Commutation Relations characterise the corresponding duality of position and momentum observables in the symmetry picture, by specifying a braiding relation between the space-translation symmetry and the momentum-boost symmetry. When time is quantised, the duality between time-translation symmetry $T_t$ and energy-shift symmetry $S_E$ in a quantum clock can similarly be characterised by the Weyl CCRs, as follows:
\begin{equation}
	\hfill
	S_E T_t = e^{i 2\pi Et } T_t S_E
	\hfill
\end{equation}
This can be equivalently expressed in terms of the adjoint of $S_E$, to match the exact formulation of our result below:
\begin{equation}
	\hfill
	S_E^\dagger T_t = e^{-i 2\pi Et } T_t S_E^\dagger
	\hfill
\end{equation}
In our formalism, the Weyl CCRs are an immediate consequence of a diagrammatic axiom of interacting quantum observables.
\begin{proposition}
	\label{proposition:weyl}
	The Weyl CCRs for time and energy duality are an immediate consequence of the bialgebra law:
	\begin{equation}
		\hfill
			\scalebox{\picturescaling}{$
				\input{figures/WeylCCRR.tikz}
				\raisebox{2cm}{\hspace{5mm}$=$\hspace{8mm}}
				\input{figures/WeylCCRM.tikz}
				\raisebox{2cm}{\hspace{7mm}$=$\hspace{5mm}}
				\input{figures/WeylCCRL.tikz}
				\hspace{-4mm}
				\input{figures/WeylCCRScalar.tikz}
			$}
		\hfill
	\end{equation}
	recalling that $\braket{E}{t} = e^{-i 2\pi Et }$.
\end{proposition}

\subsection{Stone's Theorem on 1-parameter unitary groups}

Stone's Theorem on 1-parameter unitary groups is a key result in dynamics, showing that dynamics can be uniquely reconstructed from the Hamiltonian observable. In the symmetry picture, it can be stated as follows:
\begin{equation}
	\hfill
	\alpha_t = \int_{G^\wedge} e^{-i 2\pi Et} P_E dE
	\hfill
\end{equation}
where $G$ is the time-translation group, $\alpha$ is the quantum dynamical system and $P_E$ are the projectors on the energy eigenspaces. We are working in the non-standard settings, so that the integral $\int_{G^\wedge} dE$ is really just a sum.\cite{robinson1974nonstandard}
\begin{proposition}
	\label{proposition:stone}
	Stone's Theorem is a consequence of diagrammatic time-energy duality:
	\begin{equation}
		\hfill
			\scalebox{\picturescaling}{$
				\input{figures/StoneL.tikz}
				\raisebox{2cm}{\hspace{5mm}$=$\hspace{7mm}}
				\input{figures/StoneM.tikz}
				\raisebox{2cm}{\hspace{5mm}$=$\hspace{7mm}}
				\raisebox{2cm}{$\sum_{E}$\hspace{2mm}}
				\input{figures/StoneR.tikz}
			$}
		\hfill
	\end{equation}
	recalling that $\braket{E}{t} = e^{-i 2\pi Et }$.
\end{proposition}

\subsection{von Neumann's Mean Ergodic Theorem}

In a rather precise sense, von Neumann's Mean Ergodic Theorem is the inverse of Stone's Theorem, showing that the Hamiltonian observable can be reconstructed from the dynamics. The usual formulation of von Neumann's Theorem only talks about the ground energy eigenspace, but an equivalent formulation can be used to reconstruct all energy eigenspaces:
\begin{equation}
	\hfill
	P_E = \frac{1}{|G|}\int_{G} e^{i 2\pi Et} \alpha_t dt
	\hfill
\end{equation}
where $G$ is the time-translation group, $\alpha$ is the quantum dynamical system and $P_E$ are the projectors on the energy eigenspaces. By $|G|$ we literally mean the size of $G$, which is a well-defined scalar $|G| = \omega$ in the non-standard world: physically, this is the volume $\omega = \omega_\text{uv}\omega_\text{ir}$ of time-energy configuration space. The equation above is the limit-free, non-standard equivalent of the usual formulation of von Neumann's theorem in terms of limits. As with Stone's Theorem before, the integral $\int_{G} dt$ is really just a sum $\sum_{}$ in the non-standard setting.
\begin{proposition}
	\label{proposition:vn}
	von Neumann's Mean Ergodic Theorem is a consequence of diagrammatic time-energy duality:
	\begin{equation}
		\hfill
			\scalebox{\picturescaling}{$
				\input{figures/vNL.tikz}
				\raisebox{2.9cm}{\hspace{-2mm}$\dfrac{1}{\omega}$}
				\raisebox{2cm}{\hspace{5mm}$=$\hspace{7mm}}
				\input{figures/vNM.tikz}
				\raisebox{0.9cm}{\hspace{-2mm}$\dfrac{1}{\omega}$}
				\raisebox{2cm}{\hspace{5mm}$=$\hspace{7mm}}
				\raisebox{2cm}{$\dfrac{1}{|G|}\sum_{t}$\hspace{2mm}}
				\input{figures/vNR.tikz}
			$}
		\hfill
	\end{equation}
	recalling that $\braket{t}{E} = e^{i 2\pi Et}$.
\end{proposition}

\section{Conclusions}

In this work, we have presented a diagrammatic framework to reason about quantum dynamics, using algebras and coalgebras for a monad and a comonad induced by a pair of interacting quantum observables. We have been able to treat dynamics, both discrete and continuous, of finite- and infinite-dimensional quantum systems, thanks to the rich tool-set provided by hyperfinite non-standard Hilbert spaces. We have shown that our framework yields completely straightforward diagrammatic proofs of some key results in quantum dynamics.

In future work, we will explore the foundational and computational implications of our new framework. Specifically, we will detail the applications to the problem of time observable and to the formulation of Feynman's Clock, already sketched in the author's DPhil Thesis \cite{gogioso2017thesis}.

\newpage
\nocite{*}
\bibliography{biblio}

\newpage
\appendix

\section{Categorical Quantum Mechanics}

\emph{Categorical quantum mechanics} takes its roots in the seminal work \cite{abramsky2004categorical,abramsky2009} and a detailed treatment of the first decade of work in the field can be found in the 900+ page monograph \cite{coecke_kissinger_book}. Here we recap some fundamentals of the formalism, for the benefit of readers from different communities who may be unfamiliar with them.

\subsection{Symmetric monoidal categories}

The general motivation behind the application of category-theoretic tools lies in the intuition that the features distinguishing quantum theory from classical physics can be understood in terms of the way quantum processes compose, sequentially and in parallel: this forms the basis of the \emph{process-theoretic} description of quantum theory. The mathematical arena for such process-theoretic description is that of symmetric monoidal categories:
\begin{itemize}
	\item physical systems are objects;
	\item processes between systems are morphisms;
	\item sequential composition of processes is composition of morphisms;
	\item parallel composition of processes is tensor product of morphisms;
	\item the process of doing nothing to a system is the identity morphism;
	\item the tensor product of objects is interpreted as a joint system;
	\item the tensor unit is interpreted as a trivial system.
\end{itemize}
Of special interest in categorical quantum mechanics is the symmetric monoidal category $\fHilb$ of finite-dimensional Hilbert spaces and complex linear maps between them (with the tensor product of Hilbert spaces as tensor product of objects and the Kronecker product of complex matrices as tensor product of morphisms).

Monoidal categories have a natural diagrammatic formalism---see \cite{selinger2011survey} for a comprehensive survey---in which systems/objects $A$ are depicted as wires and processes/morphisms $f: A \rightarrow B$ are depicted as boxes.
\begin{equation}
	\hfill
	\scalebox{\picturescaling}{$
		\begin{tikzpicture}
			\node (in) at (0,-1.25) {$A$};
			\node (out) at (0,+1.25) {$A$};
			\draw[-] (in) to (out);
		\end{tikzpicture}
		\hspace{5cm}
		\begin{tikzpicture}
			\node (in) at (0,-1.25) {$A$};
			\node (out) at (0,+1.25) {$B$};
			\node[box] (box) at (0,0) {$f$};
			\draw[-] (in) to (box);
			\draw[-] (box) to (out);
		\end{tikzpicture}
	$}
	\hfill
\end{equation}
Tensor product of systems is depicted by stacking the corresponding wires side-by-side horizontally. As a convention, we draw our diagrams bottom-to-top, so that the wire(s) corresponding to the domain of a morphism (the \emph{inputs} of the process) are at the bottom and the wire(s) corresponding to the codomain of a morphism (the \emph{outputs} of the process) are at the top, so that a generic morphism/process $f: A_1 \otimes...\otimes A_n \rightarrow B_1 \otimes...\otimes B_m$ is depicted as follows:
\begin{equation}
	\hfill
	\scalebox{\picturescaling}{$
		\begin{tikzpicture}
			\node (in) at (-0.5,-1.2) {$A_1$};
			\node (out) at (-0.5,+1.2) {$B_1$};
			\draw[-] (in) to (out);
			\node (in) at (0.5,-1.2) {$A_n$};
			\node (out) at (0.5,+1.2) {$B_m$};
			\draw[-] (in) to (out);
			\node (dots) at (0,-1.2) {$...$};
			\node (dots) at (0,+1.2) {$...$};
			\node[box,minimum width=2cm] (box) at (0,0) {$f$};
			\node at (-1.75,-1.2) {inputs $\rightarrow$};
			\node at (-1.8,+1.2) {outputs $\rightarrow$};
		\end{tikzpicture}
	$}
	\hfill
\end{equation}
Sequential composition $g \circ f: A \rightarrow C$ of two processes $f: A \rightarrow B$ and $g: B \rightarrow C$ is depicted by stacking the corresponding boxes vertically and connecting the output wires of $f$ to the input wires of $g$; parallel composition $f \otimes g: A \otimes C \rightarrow B \otimes D$ of two processes $f: A \rightarrow B$ and $g: C \rightarrow D$ is depicted by stacking the corresponding boxes side-by-side horizontally; the symmetry isomorphisms $A \otimes B \rightarrow B \otimes A$ are depicted by a crossing of the wires:
\begin{equation}
	\hfill
	\scalebox{\picturescaling}{$
		\begin{tikzpicture}
			\node (in) at (0,-2) {$A$};
			\node (out) at (0,+2) {$C$};
			\draw[-] (in) to (out);
			\node[box] (f) at (0,-0.8) {$f$};
			\node[box] (g) at (0,+0.8) {$g$};
		\end{tikzpicture}
	$}
	\hspace{3cm}
	\raisebox{0.6cm}{
	\scalebox{\picturescaling}{$
		\begin{tikzpicture}
			\node (in) at (-0.75,-1.2) {$A$};
			\node (out) at (-0.75,+1.2) {$B$};
			\draw[-] (in) to (out);
			\node (in) at (0.75,-1.2) {$C$};
			\node (out) at (0.75,+1.2) {$D$};
			\draw[-] (in) to (out);
			\node[box] (f) at (-0.75,0) {$f$};
			\node[box] (g) at (+0.75,0) {$g$};
		\end{tikzpicture}
	$}
	}
	\hspace{3cm}
	\raisebox{0.75cm}{
	\scalebox{\picturescaling}{$
		\begin{tikzpicture}
			\node (inl) at (-0.75,-1) {$A$};
			\node (outl) at (-0.75,+1) {$B$};
			\node (inr) at (0.75,-1) {$B$};
			\node (outr) at (0.75,+1) {$A$};
			\draw[-,out=90,in=270] (inl) to (outr);
			\draw[-,out=90,in=270] (inr) to (outl);
		\end{tikzpicture}
	$}
	}
	\hfill
\end{equation}
Two diagrams are considered equal if they are equal up to isotopy, keeping the (relative ordering of the) input and output endpoints fixed: this principle is often referred to as ``only topology matters''. For example, a special case of bifunctoriality for the tensor product holds by ``sliding'' the two boxes vertically (on the left), while naturality of the symmetry isomorphism is obtained by ``sliding boxes through each other'' over a wire crossing (on the right):
\begin{equation}
	\hfill
	\scalebox{\picturescalingmore}{$
		\begin{tikzpicture}
			\node (in) at (-0.75,-2) {};
			\node (out) at (-0.75,+2) {};
			\draw[-] (in) to (out);
			\node (in) at (0.75,-2) {};
			\node (out) at (0.75,+2) {};
			\draw[-] (in) to (out);
			\node[box] (f) at (-0.75,0.75) {$f$};
			\node[box] (g) at (+0.75,-0.75) {$g$};
		\end{tikzpicture}
	\raisebox{2cm}{
	\hspace{1cm} = \hspace{1cm}
	}
		\begin{tikzpicture}
			\node (in) at (-0.75,-2) {};
			\node (out) at (-0.75,+2) {};
			\draw[-] (in) to (out);
			\node (in) at (0.75,-2) {};
			\node (out) at (0.75,+2) {};
			\draw[-] (in) to (out);
			\node[box] (f) at (-0.75,-0.75) {$f$};
			\node[box] (g) at (+0.75,0.75) {$g$};
		\end{tikzpicture}
	$}
	\hspace{2cm}
	\raisebox{4mm}{
	\scalebox{\picturescalingmore}{$
		\begin{tikzpicture}
			\node (inl) at (-0.75,-1) {};
			\node[box] (outl) at (-0.75,+1) {$f$};
			\node (outlout) at (-0.75,+2) {};
			\node (inr) at (0.75,-1) {};
			\node[box] (outr) at (0.75,+1) {$g$};
			\node (outrout) at (0.75,+2) {};
			\draw[-,out=90,in=270] (inl) to (outr);
			\draw[-,out=90,in=270] (inr) to (outl);
			\draw[-] (outl) to (outlout);
			\draw[-] (outr) to (outrout);
		\end{tikzpicture}
		\raisebox{1.5cm}{
		\hspace{1cm} = \hspace{1cm}
		}
		\begin{tikzpicture}
			\node[box] (inl) at (-0.75,-1) {$g$};
			\node (outl) at (-0.75,+1) {};
			\node (inlin) at (-0.75,-2) {};
			\node[box] (inr) at (0.75,-1) {$f$};
			\node (outr) at (0.75,+1) {};
			\node (inrin) at (0.75,-2) {};
			\draw[-,out=90,in=270] (inl) to (outr);
			\draw[-,out=90,in=270] (inr) to (outl);
			\draw[-] (inlin) to (inl);
			\draw[-] (inrin) to (inr);
		\end{tikzpicture}
	$}
	}
	\hfill
\end{equation}
Planar isotopy is sufficient for monoidal categories. For symmetric monoidal categories, on the other hand, a little amount of 4d space is used for sliding across wire crossings.

\subsection{States, scalars and effects}

Special cases of boxes are those without any input and/or any output wires:
\begin{equation}
	\hfill
	\scalebox{\picturescaling}{$
		\begin{tikzpicture}
			\node (out) at (0,+1.5) {};
			\node[box] (box) at (0,0) {$\psi$};
			\draw[-] (box) to (out);
			\node at (0,-1.8) {no inputs};
		\end{tikzpicture}
	$}
	\hspace{3cm}
	\scalebox{\picturescaling}{$
		\begin{tikzpicture}
			\node (in) at (0,-1.5) {};
			\node[box] (box) at (0,0) {$a$};
			\draw[-] (in) to (box);
			\node at (0,-1.8) {no outputs};
		\end{tikzpicture}
	$}
	\hspace{3cm}
	\scalebox{\picturescaling}{$
		\begin{tikzpicture}
			\node (box) at (0,0) {$\xi$};
			\node at (0,-1.8) {no inputs/outputs};
		\end{tikzpicture}
	$}
	\hfill
\end{equation}
Boxes with no input wires are called \emph{states} and they correspond to the process of producing something in a system starting from nothing (aka the trivial system). In the category $\fHilb$, states correspond exactly to vectors in a Hilbert space, i.e. to kets $\ket{\psi}$. Boxes with no inputs nor outputs are called \emph{scalars}. In the category $\fHilb$, scalars correspond to complex numbers $\xi$ and we will write them as floating numbers. Finally, boxes with no outputs are called \emph{effects}. In the category $\fHilb$, effects correspond exactly to covectors in a Hilbert space, i.e. to bras $\bra{a}$. Effects can be though of as the process of conditioning on the outcome of a (non-degenerate demolition) quantum measurement: applied to a state in a system, an effect returns a complex number (the norm squared of which yields the outcome probability, according to the Born rule).

\subsection{Dagger symmetric monoidal categories}

The symmetric monoidal categories of interest in categorical quantum mechanics are equipped with an involutive op-functor, the \emph{dagger}, which sends morphisms $f: A \rightarrow B$ to morphisms $f^\dagger: B \rightarrow A$. In the diagrammatic formalism, the dagger is depicted as a vertical mirror symmetry:
\begin{equation}
	\hfill
	\scalebox{\picturescaling}{$
		\begin{tikzpicture}
			\node (in) at (-0.5,-1.2) {$A_1$};
			\node (out) at (-0.5,+1.2) {$B_1$};
			\draw[-] (in) to (out);
			\node (in) at (0.5,-1.2) {$A_n$};
			\node (out) at (0.5,+1.2) {$B_m$};
			\draw[-] (in) to (out);
			\node (dots) at (0,-1.2) {$...$};
			\node (dots) at (0,+1.2) {$...$};
			\node[box,minimum width=2cm] (box) at (0,0) {$f$};
		\end{tikzpicture}
	$}
	\raisebox{1cm}{
	\hspace{1cm} $\mapsto$ \hspace{1cm}
	}
	\scalebox{\picturescaling}{$
		\begin{tikzpicture}
			\node (in) at (-0.5,-1.2) {$B_1$};
			\node (out) at (-0.5,+1.2) {$A_1$};
			\draw[-] (in) to (out);
			\node (in) at (0.5,-1.2) {$B_m$};
			\node (out) at (0.5,+1.2) {$A_n$};
			\draw[-] (in) to (out);
			\node (dots) at (0,-1.2) {$...$};
			\node (dots) at (0,+1.2) {$...$};
			\node[box,minimum width=2cm] (box) at (0,0) {$f^\dagger$};
		\end{tikzpicture}
	$}
	\hfill
\end{equation}
In the dagger symmetric monoidal category $\fHilb$, the dagger is (chosen to be) the operation of taking the adjoint (i.e. the conjugate transpose of matrices). In particular, the dagger sends a state $\ket{\psi}$ to the corresponding effect $\bra{\psi}$ and vice-versa. On scalars, the dagger acts as complex conjugation $\xi^\dagger = \xi^\ast$.

\subsection{Quantum observables}

Some boxes have special significance and a special notation is reserved to them. The most important case is that of \emph{symmetric special $\dagger$-Frobenius algebras} $\raisebox{-0.5pt}{\hbox{\input{symbols/ZdotSym.tex}}}\!\! := (\mathcal{H},\raisebox{-2.5pt}{\hbox{\input{symbols/ZmultSym.tex}}}\!,\!\raisebox{-2.5pt}{\hbox{\input{symbols/ZunitSym.tex}}}\!\!,\raisebox{-2.5pt}{\hbox{\input{symbols/ZcomultSym.tex}}}\!,\!\raisebox{-2.5pt}{\hbox{\input{symbols/ZcounitSym.tex}}}\!\!)$: these are depicted by coloured dots with wires connected to them---lovingly referred to as \emph{spiders} in the literature---and the axioms defining them are given in the main text.

The reason why special symmetric $\dagger$-Frobenius algebras are of key interest in categorical quantum mechanics is their correspondence in the category $\fHilb$ to quantum observables, i.e. to finite-dimensional C*-algebras \cite{vicary2011categorical}. In particular, \emph{commutative} special $\dagger$-Frobenius algebras---often referred to as \emph{classical structures} in the literature---correspond bijectively to orthonormal bases, where the basis vectors $\ket{g}$ in each basis are given by the \emph{classical states} for the algebra, i.e. those satisfying the Equation \eqref{eqn_classicalStates}. For a given orthonormal basis $(\ket{g})_{g}$, the comultiplication is obtained as $\raisebox{-2.5pt}{\hbox{\input{symbols/ZcomultSym.tex}}}\! := \sum_{g} (\ket{g}\otimes\ket{g})\bra{g}$, i.e. the map $\ket{g} \mapsto \ket{g} \otimes \ket{g},$ the counit as $\!\raisebox{-2.5pt}{\hbox{\input{symbols/ZcounitSym.tex}}}\!\! := \sum_{g} \bra{g}$, i.e. the map $\ket{g} \mapsto 1$, the multiplication $\raisebox{-2.5pt}{\hbox{\input{symbols/ZmultSym.tex}}}\!$ and unit $\!\raisebox{-2.5pt}{\hbox{\input{symbols/ZunitSym.tex}}}\!\!$ as their adjoints. More generally, the orthogonal projectors $p:\mathcal{H} \rightarrow \mathcal{H}$ in a quantum observable can be characterised as the central, self-adjoint, idempotent elements for the algebra.\footnote{See \cite{vicary2011categorical} for the full proof and Section 2.4.2 of the author's DPhil thesis \cite{gogioso2017thesis} for a summary, noting that a left-to-right diagrammatic convention is adopted in the latter.}

\subsection{Dagger compact structure}

In the category $\fHilb$, each system $\mathcal{H}$ is equipped with a classical structure $\raisebox{-0.5pt}{\hbox{\input{symbols/ZdotSym.tex}}}\!\!$ for each choice of orthonormal basis. Each such classical structure---and more generally each symmetric special $\dagger$-Frobenius algebras on $\mathcal{H}$---induces a self-duality on $\mathcal{H}$ through its cup and cap, because of the snake equation \eqref{eqn_snakeEqn}, which holds by planar isotopy. The cup and cap are symmetric and related by the dagger, making $\fHilb$ a \emph{dagger compact category}.

\subsection{Infinite-dimensional categorical quantum mechanics}

The main obstacle to the extension of categorical quantum mechanics to infinite dimensions is the disappearance of symmetric special $\dagger$-Frobenius algebras: while it is true that for a complete orthonormal basis one can still define the comultiplication $\raisebox{-2.5pt}{\hbox{\input{symbols/ZcomultSym.tex}}}\! = \sum_{n=1}^{\infty} (\ket{n} \otimes \ket{n}) \bra{n}$ and multiplication $\raisebox{-2.5pt}{\hbox{\input{symbols/ZmultSym.tex}}}\!$ \cite{abramsky2012hstar}, the unit/counit  would give rise to infinite-norm states $\!\raisebox{-2.5pt}{\hbox{\input{symbols/ZunitSym.tex}}}\!\! = \sum_{n=1}^{\infty} \ket{n}$. While \cite{abramsky2012hstar} suggests it may be possible to overcome the absence of units in this context, the non-existence of infinite group algebras---necessary to this work---is intrinsically related to the presence of infinities and cannot be fixed directly.

Enter non-standard analysis. Because \emph{approximate} units $\!\raisebox{-2.5pt}{\hbox{\input{symbols/ZunitSym.tex}}}\!\!^{(\nu)} = \sum_{n=1}^{\nu} \ket{n}$ for the algebra associated to a basis exist for all $\nu \in \naturals$, by Transfer Theorem \cite{robinson1974nonstandard,goldblatt1998hyperreals} we can take $\nu$ to be some \emph{infinite} natural number and obtain a genuine unit $\!\raisebox{-2.5pt}{\hbox{\input{symbols/ZunitSym.tex}}}\!\! = \sum_{n=1}^{\nu} \ket{n}$, multiplication, counit and comultiplication $\raisebox{-2.5pt}{\hbox{\input{symbols/ZcomultSym.tex}}}\! = \sum_{n=1}^{\nu} (\ket{n} \otimes \ket{n}) \bra{n}$ for a special commutative $\dagger$-Frobenius algebra. Moreover, by Transfer Theorem we can also formulate group algebras for all abelian groups with $\nu$ elements. The full details can be found in \cite{gogioso2017infinite,gogioso2018qft} and in Section 3.5 of the author's DPhil thesis \cite{gogioso2017thesis}.

TL;DR: we work in the dagger compact category $\starHilb$ of non-standard Hilbert spaces with dimension some non-standard natural number $\nu \in \starNaturals$: from the non-standard perspective these spaces are finite-dimensional, so the Transfer Theorem can be used to lift many of the structures and properties of the dagger compact category $\fHilb$. In particular, the algebraic manipulation of vectors, matrices and scalars in $\starHilb$ is analogous to that of their $\fHilb$ counterparts.
When related back to standard Hilbert spaces, however, the objects of $\starHilb$ cover much more than $\fHilb$, including both the separable infinite-dimensional Hilbert spaces used in traditional quantum mechanics and the non-separable ones arising in quantum field theory.

\section{Proofs}

\subsection*{Proof of Proposition \ref{proposition:dagger-frobenius-monad-transpose-condition}}
\begin{proof}
	Substitute the antipode for its definition and apply the snake equation for $\raisebox{-0.5pt}{\hbox{\input{symbols/ZdotSym.tex}}}\!\!$.
\end{proof}

\subsection*{Proof of Proposition \ref{proposition:non-standard-time-translation-groups}}
\begin{proof}
	If we take $\omega_\text{uv}$ finite and $\omega_\text{ir}$ finite, then $\omega = N \in \naturals$, all elements in $\frac{1}{\omega_\text{uv}}\starIntegersMod{\omega}$ are finite and taking the standard part yields $\frac{1}{\stdpart{\omega_\text{uv}}}\integersMod{\omega}$.
	If we take $\omega_\text{uv}$ finite and $\omega_\text{ir}$ infinite, then the finite elements in $\frac{1}{\omega_\text{uv}}\starIntegersMod{\omega}$ are those in the form $\frac{1}{\omega_\text{uv}}\integers$ and taking the standard part yields $\frac{1}{\stdpart{\omega_\text{uv}}}\integers.$
	If we take $\omega_\text{uv}$ infinite and $\omega_\text{ir}$ finite, then we have the following subgroup inclusion
	\begin{equation}
		\hfill
		\frac{1}{\omega_\text{uv}}\starIntegersMod{\omega}
		=
		\frac{\omega_\text{ir}}{\omega}\starIntegersMod{\omega}
		=
		\suchthat{\frac{n\omega_\text{ir}}{\omega}}{n \in \starIntegersMod{\omega}}
		<
		\starReals/\omega_\text{ir}\starIntegers
		\hfill
	\end{equation}
	All elements are finite and taking the standard part yields $\reals/\stdpart{\omega_\text{ir}}\integers$.
	If we take $\omega_\text{ir}$ infinite and $\omega_\text{uv}$ infinite, finally, we have the following subset inclusion:
	\begin{equation}
		\hfill
		\frac{1}{\omega_\text{uv}}\starIntegersMod{\omega}
		=
		\suchthat{\frac{n}{\omega_\text{uv}}}{n \in \left\{-\left\lfloor \frac{\omega-1}{2} \right\rfloor,...,+\left\lfloor \frac{\omega}{2} \right\rfloor\right\}}
		\subset
		\starReals
		\hfill
	\end{equation}
	The finite elements cover the finite elements of $\starReals$ with infinitesimal mesh, hence taking the standard part yields $\reals$.
\end{proof}

\subsection*{Proof of Proposition \ref{proposition:quantum-clocks-existence}}
\begin{proof}
	Because $\omega := \omega_\text{uv} \omega_\text{ir} \in \starNaturals$, by the Transfer Principle we always have an object $\mathcal{G}$ of $\starHilb$ with orthonormal basis $(\ket{e_i})_{i \in \starIntegersMod{\omega}}$. Let $\raisebox{-0.5pt}{\hbox{\input{symbols/ZdotSym.tex}}}\!\!$ be the special commutative $\dagger$-Frobenius algebra associated to the orthonormal basis, define $\raisebox{-2.5pt}{\hbox{\input{symbols/XunitSym.tex}}}\!$ to be $\ket{0}$ and $\raisebox{-2.5pt}{\hbox{\input{symbols/XmultSym.tex}}}\!$ to be the linear extension of the multiplication in $\frac{1}{\omega_\text{uv}}\starIntegersMod{\omega}$. Then $(\mathcal{G},\raisebox{-0.5pt}{\hbox{\input{symbols/ZdotSym.tex}}}\!\!,\raisebox{-0.5pt}{\hbox{\input{symbols/XdotSym.tex}}}\!\!)$ is a pair of interacting quantum observables in $\starHilb$ corresponding to the group algebra $\starComplexs[\frac{1}{\omega_\text{uv}}\starIntegersMod{\omega}]$, as we wanted.
\end{proof}

\subsection*{Proof of Proposition \ref{proposition:pontryagin-duality}}
\begin{proof}
	The possible values of energy $E$ must correspond bijectively with the possible unitary group homomorphisms $G \rightarrow \complexs$ yielding the phases acquired under time-translation by energy eigenstates. Canonically, such homomorphisms are the elements of the Pontryagin dual $G^\wedge$.

	Checking that the plane-waves $\ket{E}$ are the classical states for $\raisebox{-0.5pt}{\hbox{\input{symbols/XdotSym.tex}}}\!\!$ is straightforward. Because clock time states form an orthonormal basis, we can biject the effects $\bra{E}$ with the multiplicative characters $\chi_E: t \mapsto e^{-i2\pi Et} \in G^\wedge$. The (adjoint of the) copy condition is multiplicativity of characters $\chi_E(t+s) = \chi_E(t) \chi_E(s)$. The (adjoint of the) delete condition is the condition that $\chi_E(0) = 1$. The (adjoint of the) self-conjugacy condition, finally, is unitarity of characters $\chi_E(t)^\dagger = \chi_E(-t)$.

	Under the identification of $\bra{E}$ with $\chi_E$, it is immediate to see that $\raisebox{-2.5pt}{\hbox{\input{symbols/ZmultSym.tex}}}\!$ acts as pointwise multiplication of characters and that $\!\raisebox{-2.5pt}{\hbox{\input{symbols/ZunitSym.tex}}}\!\!$ corresponds to the trivial character, so that $G^\wedge$ is obtained by taking the standard part of the finite elements in $\frac{1}{\omega_\text{ir}}\starIntegersMod{\omega}$.
\end{proof}

\subsection*{Proof of Proposition \ref{proposition:algebras-dynamical-systems}}
\begin{proof}
	By requiring $\alpha_t$ to be near-standard for all $t$ we have singled out exactly those representations of the non-standard group which yield representations $\stdpart{\alpha_{\stdpart{t}}}$ for the corresponding standard group. The defining equations for unitary representations are already satisfied. The defining equation for a morphism $\Phi: \alpha \rightarrow \beta$ implies that $\Phi \alpha_t = \beta_t \Phi$ for all $t$, so that morphism of algebras give equivariant maps of representations. Synchronised parallel composition speaks for itself.
\end{proof}

\subsection*{Proof of Proposition \ref{proposition:histories}}
\begin{proof}
	The proof is entirely by straightforward diagrammatic manipulation, based on the observation that $\Psi$ is exactly the time evolution of the state $\Psi(0)$:
	\begin{equation}
		\hfill
			\scalebox{\picturescalingmore}{$
				\input{figures/historyProof1L.tikz}
				\raisebox{2cm}{\hspace{5mm}$=$\hspace{5mm}}
				\input{figures/historyProof1M.tikz}
				\raisebox{2cm}{\hspace{5mm}$=$\hspace{5mm}}
				\input{figures/historyProof1R.tikz}
			$}
		\hfill
	\end{equation}
\end{proof}

\subsection*{Proof of Proposition \ref{proposition:hamiltonian}}
\begin{proof}
	That $\alpha^\dagger$ is a coalgebra for $\_ \otimes \raisebox{-0.5pt}{\hbox{\input{symbols/ZdotSym.tex}}}\!\!$ is a straightforward diagrammatic check using the algebra equations for $\alpha$ and the snake equations. The fact that Schr\"{o}dinger's Equation holds for the eigenstates of projectors is another straightforward diagrammatic check.
\end{proof}

\subsection*{Proof of Propositions \ref{proposition:weyl}, \ref{proposition:stone} and \ref{proposition:vn}}

The proofs are already essentially in the respective diagrammatic statements.

\end{document}